\documentclass[journal]{IEEEtran}

\usepackage{cite}
\usepackage{algorithm,algorithmicx}
\usepackage{algpseudocode}
\usepackage[small,bf]{caption}
\usepackage[cmex10]{amsmath}
\usepackage{amssymb,latexsym,epsfig,subfigure,epic,amscd,mathrsfs,euscript,bm}
\usepackage{color}
\usepackage{booktabs}
\usepackage{array}

\usepackage{amsfonts}
\usepackage{amsopn}
\usepackage{graphicx}
\usepackage{url} 
\usepackage{empheq}

\newtheorem{myprop}{\bf{Proposition}}
\newtheorem{theorem}{\bf{Theorem}}
\newtheorem{remark}{\bf{Remark}}
\newtheorem{lemma}{\bf{Lemma}}

\DeclareMathOperator{\tr}{tr}

\DeclareMathOperator{\cov}{cov}

\DeclareMathOperator*{\minimize}{\text{minimize}}
\DeclareMathOperator*{\maximize}{\text{maximize}}

\DeclareMathOperator*{\st}{\text{subject to}}
\DeclareMathAlphabet\mathbfcal{OMS}{cmsy}{b}{n}
\newcommand{\Def}[0]{\mathrel{\mathop:}=}

\begin{document}
 
\title{Optimal  Sensor Collaboration  for Parameter Tracking Using Energy Harvesting Sensors}

\author{Shan~Zhang, 
        Sijia~Liu,~\IEEEmembership{Member,~IEEE,}
         Vinod~Sharma,~\IEEEmembership{Senior Member,~IEEE,} 
	 and Pramod~K.~Varshney,~\IEEEmembership{Life Fellow,~IEEE}	
\thanks{Copyright (c) 2015 IEEE. Personal use of this material is permitted. However, permission to use this material for any other purposes must be obtained from the IEEE by sending a request to pubs-permissions@ieee.org.}
\thanks{S. Zhang and P. K. Varshney are with the Department
of EECS, Syracuse University, Syracuse, NY 13244, USA. Email:  \{szhang60,varshney\}@syr.edu.}
\thanks{S. Liu  is with the MIT-IBM Watson AI Lab, IBM Research, Cambridge, MA 02142, USA. Email: sijia.liu@ibm.com.}
\thanks{V. Sharma is with the Department of ECE, Indian Institute of Science, Bangalore 560012, India. Email: vinod@ece.iisc.ernet.in.}
\thanks{
This work was supported in part by ARO grant number W911NF-14-1-0339 and AFOSR grant number FA9550-16-1-0077.}
}

\maketitle

\begin{abstract}
In this paper, we design an optimal sensor collaboration strategy among neighboring nodes while tracking a time-varying parameter using wireless sensor networks in the presence of imperfect communication channels. The sensor network is assumed to be self-powered, where sensors are equipped with energy harvesters that  replenish energy from the environment. 
In order to minimize the mean square estimation error of parameter tracking,
we propose an online sensor collaboration policy   subject to  real-time energy harvesting constraints. The proposed energy allocation strategy is computationally light and only relies on the second-order statistics of the system parameters. For this, we first consider 
an offline non-convex optimization problem, which is  solved exactly 
using  semidefinite programming.  
Based on the offline solution, we design an online power allocation policy that requires minimal
online computation and satisfies the dynamics of energy flow at each
sensor. We prove that the proposed online policy is asymptotically equivalent to the optimal offline solution and show its convergence rate and robustness.
We empirically show that the estimation performance of the proposed online scheme 
is better than that of the online scheme when channel state information about the dynamical system is available in the low SNR regime.  Numerical results are conducted to demonstrate the effectiveness of our approach.
\end{abstract}


\section{Introduction}
Recent advances in wireless communications and electronics have enabled the development of low-cost, low-power, multifunctional sensor nodes that are deployed as networks for many applications such as environment monitoring, source localization and target tracking \cite{yick2008wireless, cheng2012survey, he2006achieving}. These sensor nodes sense and measure some attributes of the targets of interest, and are able to exchange their information through in-network communications \cite{olfati2007consensus}.
In this paper, we study the problem of parameter tracking in the presence of inter-sensor communications  
that is referred to as sensor collaboration. Here sensors are   equipped with energy harvesters, which
can replenish energy from the environment, such as sun light and wind.
The act of sensor collaboration allows 
sensors  to update their measurements by taking a linear combination of the measurements of those they interact with, prior to transmission to a fusion center (FC). As suggested in 
\cite{kar2013linear}, 
 {collaboration} smooths out the observation noise,
thereby improving the quality of the signal transmitted to the FC and eventual estimation performance.


In the absence of sensor collaboration, the proposed estimation architecture reduces to a classical distributed estimation system that uses an  amplify-and-forward transmission strategy \cite{cuixiagolluopoo07, xiao2008linear, knorn2015distortion}. The power allocation problem for distributed estimation under an \textit{orthogonal} multiple access channel (MAC) model was studied in \cite{cuixiagolluopoo07}, where
 the optimal power amplifying factors were found  by minimizing the estimation distortion  subject to   energy constraints. In \cite{xiao2008linear}, the same problem was studied under a different communication model, \textit{coherent} MAC, where sensors coherently form a beam into a common channel received at the FC. In \cite{knorn2015distortion}, the design of optimal power allocation strategies was addressed over orthogonal fading wireless channels for  source estimation in a network of energy-harvesting sensors.
 In \cite{leong2011asymptotics, jiang2014optimal}, the problem of power allocation for parameter tracking was studied where the (scalar) parameter of interest was modeled as a  Gauss-Markov process. In the aforementioned literature \cite{cuixiagolluopoo07, xiao2008linear,knorn2015distortion,leong2011asymptotics, jiang2014optimal}, inter-sensor communication was not considered. Moreover, the work in\cite{cuixiagolluopoo07, xiao2008linear,knorn2015distortion} only focused on one-snapshot estimation, while the work \cite{leong2011asymptotics, jiang2014optimal} only studied the conventional sensor network with battery-limited sensors.
In contrast, here we seek an optimal \textit{sensor collaboration} scheme for a dynamic parameter \textit{tracking} using 
 \textit{energy-harvesting} sensors.

The problem of distributed estimation with sensor
collaboration has attracted recent attention \cite{fang2009power,thatte2008sensor,kar2013linear,liu2014sparsity, liu2015optimal,liu2016optimized}. In \cite{fang2009power}, the network topology  was assumed to be fully connected, where  all the sensors are allowed to share their observations. It was shown that the optimal strategy is to transmit the processed signal (after collaboration) over the best available channels with power levels consistent with the channel qualities. 
In \cite{thatte2008sensor}, the optimal collaboration strategy was designed under star, branch and linear network topologies. In \cite{kar2013linear,liu2014sparsity, liu2015optimal},  
the problem of collaboration network topology design was studied by incorporating the  cost of sensor collaboration. 
It was shown that a partially connected network can yield estimation performance close to that of a fully connected network. 
However, the aforementioned literature on collaborative estimation focused only on 
static networks, and the parameter to be estimated is temporally invariant.
In \cite{liu2016optimized},
the  sensor collaboration strategy was designed for the estimation of temporally correlated parameters. However, in \cite{liu2016optimized}   a batch-based linear estimator was used, where     all the historical   measurements have to be stored for decision making. 
Different from \cite{liu2016optimized}, a Kalman filter-like estimator is adopted in this paper, and a computationally inexpensive but efficient real-time sensor collaboration scheme is developed in conjunction with parameter tracking.




In the existing work on power allocation for collaborative estimation \cite{fang2009power,thatte2008sensor,kar2013linear,liu2014sparsity, liu2015optimal, liu2016optimized}, sensor networks were composed of  (conventional) battery-limited sensors. 
Motivated by the rapid advances in the emerging  
 energy-harvesting device technology \cite{priya2009energy},
it is attractive to
perform collaborative estimation using energy harvesting sensors.
In this context, each sensor 
can replenish energy from the environment  without the need of battery replacement. 


The effect of energy harvesting on parameter estimation was considered recently in
\cite{zhao2013optimal, huang2013power, nayyar2013optimal,liu2016optimal}. In \cite{zhao2013optimal}, a single sensor equipped with an energy harvester was used for parameter estimation, and the optimal power allocation scheme was designed under
  causal and non-causal side information of energy harvesting. In \cite{huang2013power}, a deterministic energy harvesting model was proposed for  parameter estimation  over an orthogonal MAC. In \cite{nayyar2013optimal}, the optimal communication (sensors to the FC) strategy was obtained for remote estimation   with energy harvesting sensors.
In \cite{zhao2013optimal, huang2013power, nayyar2013optimal}, the power allocation policy  was designed in the absence of sensor collaboration. In \cite{liu2016optimal}, the problem of energy allocation and storage control for collaborative estimation   was addressed by solving  nonconvex optimization problems through convex relaxations. However, the strategy developed in  \cite{liu2016optimal} only provides an offline solution for one snapshot estimation. In contrast, 
we propose a sensor collaboration   policy that obeys the real-time energy harvesting constraints.

Another difference between our work and 
the previous ones 
is that
the previous studies
\cite{fang2009power,thatte2008sensor,kar2013linear,liu2014sparsity, liu2015optimal, liu2016optimal, liu2016optimized} 
hinge on the noise-free inter-sensor communication model, while the noiseless assumption is relaxed in this paper. Moreover, sensors communicate with neighbors using point to point orthogonal channels in \cite{fang2009power,thatte2008sensor,kar2013linear,liu2014sparsity, liu2015optimal, liu2016optimal, liu2016optimized},
while sensors exchange their information with neighbors through a coherent MAC in this paper. Multiple-access channel communication introduces collaboration noise, which 
has to be taken into account while designing the optimal sensor collaboration scheme.


In a preliminary version of this paper \cite{liutowards},
we studied the sensor collaboration problem for static parameter estimation in the absence of collaboration noise.   
In this paper, we focus on tracking of a scalar dynamic parameter with noise-corrupted sensor collaboration. The assumption of scalar random process enables us to obtain explicit expressions for the estimation distortion and the transmission cost, and to explore the asymptotic behavior of the estimation error with respect to the energy constraint. The assumption of scalar random process was also used in other power allocation problems \cite{kar2013linear, cuixiagolluopoo07, jiang2014optimal, fang2009power}. However, different from \cite{kar2013linear, cuixiagolluopoo07, jiang2014optimal, fang2009power}, we consider a different problem in which a scalar random process is tracked in the presence of noise-corrupted inter-sensor collaboration and energy-harvesting sensors. More specifically, we have the following new contributions in this paper 
\begin{itemize}
\item We consider tracking of a correlated parameter sequence.
\item We incorporate a recursive linear minimum mean squared error estimator (R-LMMSE) to track the parameter in the presence of multiplicative noise.
\item We consider noise-corrupted sensor collaboration accomplished through a coherent MAC in a network of energy harvesting sensors for parameter tracking.
\item We propose an optimal offline sensor collaboration scheme under the assumption that the energy constraint is on an average basis, and show that although the resulting optimization problem is non-convex, a globally optimal solution can be found via semidefinite programming. Also, we prove the stability of this scheme.
\item We propose an online sensor collaboration policy that satisfies the real-time energy harvesting constraints with minimal online computational complexity, and prove that its resulting long-term estimation performance is the same as that of the optimal offline solution. Moreover, we show the convergence rate and robustness of the proposed optimal online scheme. 
\item We demonstrate the efficiency of our methodology by comparing it with a greedy optimal online policy when the channel state information (CSI) regarding the dynamical system is assumed to be known causally at the FC.
\end{itemize}

The main generalization compared to \cite{liutowards} is that the parameter to be estimated in this paper is a correlated sequence compared to the independent and identically distributed (i.i.d.) sequence in \cite{liutowards}. Thus in \cite{liutowards}, the estimation procedure required estimating the parameter via only the data obtained at that time. In this paper, we incorporate an R-LMMSE to track and estimate the parameter because of the correlation in time. Moreover, it becomes a distributed R-LMMSE with online constraints (due to energy harvesting) and multiplication of the system parameters with random coefficients (due to channel gains).  Thus, technically it is much more complicated and different from \cite{liutowards}. The optimization problem solved also is much more complicated than in \cite{liutowards}.

The rest of the paper is organized as follows. In Section\,\ref{sec: PS}, we introduce the collaborative tracking system, and present the general statement of the sensor collaboration problem. In Section\,\ref{sec: tracking}, we formulate the offline sensor collaboration problem for parameter tracking, where only partial knowledge of the second-order statistics of   system parameters is required. 
In Section\,\ref{sec: solution}, we solve the non-convex offline sensor collaboration problem and obtain its globally optimal solution via semidefinite programming. We also design an  online sensor collaboration policy. In Section\ \ref{sec: comparison}, for comparison, we study a sensor collaboration problem with known channel state information. In Section\,\ref{sec: NR}, we demonstrate the effectiveness of our online solution obtained in Section\,\ref{sec: solution} through numerical examples and compare with the greedy optimal solution obtained in Section\,\ref{sec: comparison}. Finally, in Section\ \ref{sec: conclusion} we summarize our work and discuss future research directions.

\section{Problem Statement}
\label{sec: PS}
\begin{figure}[ht!]
\centering
\includegraphics[width=7.5cm]{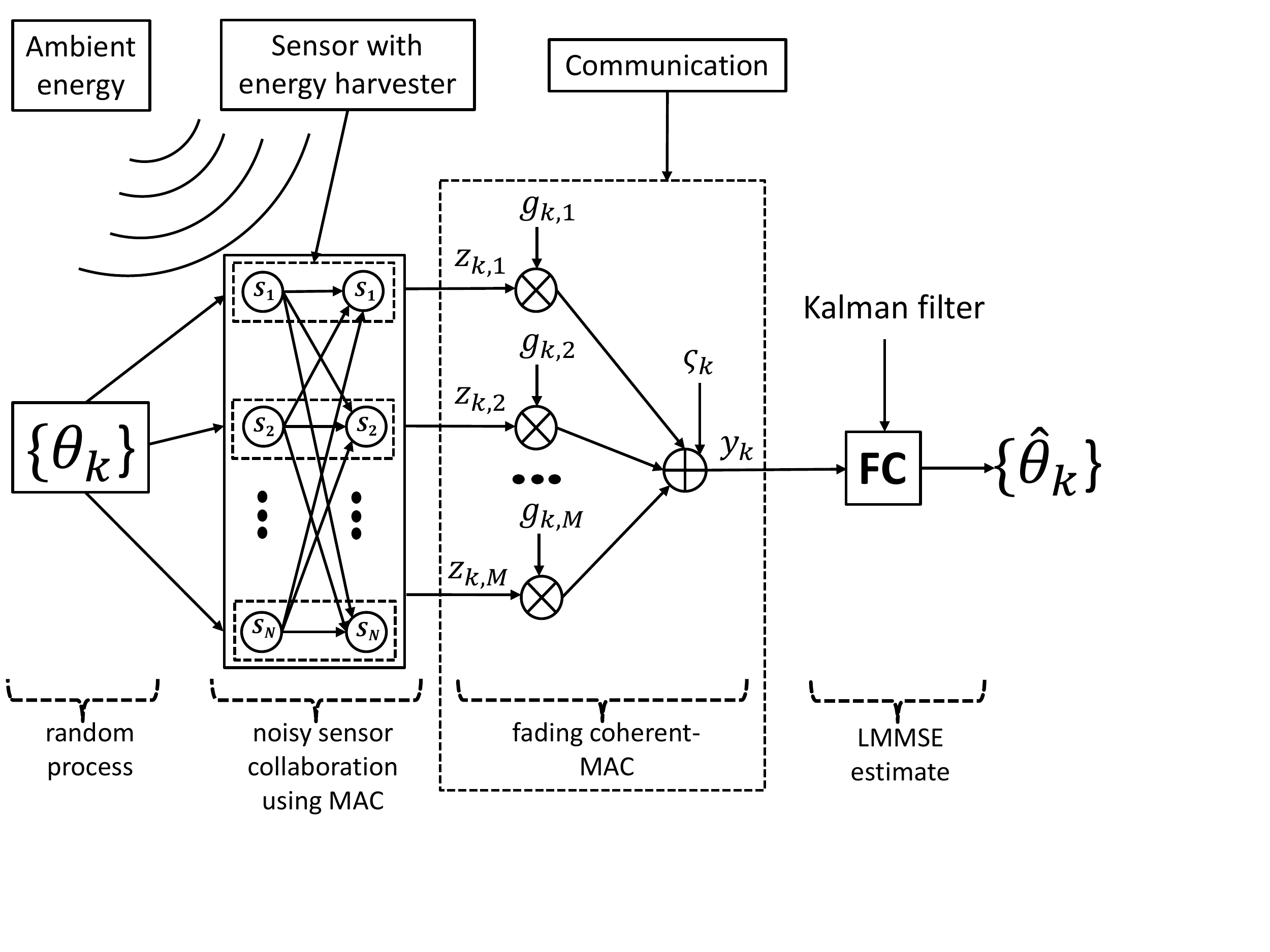} 
\caption{\footnotesize{Collaborative tracking system.}}
\label{fig: Inf_prob_MComp}
\end{figure}

In this section, we formally state the problem of sensor collaboration for parameter tracking using a network of energy harvesting sensors.
In the wireless sensor network under consideration, 
each sensor is equipped with an energy harvesting device which   replenishes itself from a renewable energy source. 
In the overall inference system,
sensors first acquire raw measurements of the phenomenon of interest via a linear sensing
model, and then perform spatial collaboration via a coherent MAC. A coherent MAC channel can be realized through transmit beamforming \cite{mudumbai2009distributed}, where sensor nodes simultaneously transmit a common message and the phases of their transmissions are controlled so that the signals constructively combine at the FC. This requires that each of the sensor nodes knows its channel gains. It has been shown in \cite{leong2011asymptotics} that the coherent MAC can be realized by the distributed synchronization schemes described under moderate amounts of phase error \cite{li2007distributed, bucklew2008convergence}. An advantage of a coherent MAC is that all the sensors use the same channel to transmit their observations. We also remark that the coherent MAC model has been widely used in the study of the power allocation problem for distributed estimation \cite{xiao2008linear, kar2013linear, jiang2014optimal, liu2016optimized, liutowards}. 
 After collaboration, the signals
are transmitted  to the FC,
which tracks  the random process. 
We show the collaborative estimation system to be studied in Fig.\,\ref{fig: Inf_prob_MComp}. A list of the system parameters is provided in Table \ref{table:list}.

\subsection{Sensor collaboration via coherent MAC}
Let $\{ \theta_k \in \mathbb R \}$ be the discrete-time random process to be estimated, which evolves as a first-order Gauss-Markov process
\begin{align}
\theta_k = \alpha \theta_{k-1} + \tau_{k}, \, k =1, 2, \ldots,
\label{eq: theta_k}
\end{align} 
where $\theta_k$ denotes the random parameter at time $k$, $\alpha$ is the known state transition coefficient,  and $\tau_{k}$ is i.i.d with zero mean and variance $\sigma_\tau^2$. 
We also assume that $\theta_k$ has zero mean, and $|\alpha| < 1$. Then, $\theta_k$ is an ergodic Markov chain with a unique stationary distribution. In addition, if $\tau_{k}$ has a positive density on whole of real line and $\mathbb E[| \tau_{k} |] < \infty$, then $\theta_k$ is geometrically ergodic \cite{meyn2012markov}. Many practically relevant distributions satisfy these conditions: Gaussian, mixtures of Gaussians and some heavy tailed stable distributions modeling electromagnetic interference.

We consider a linear measurement model
\begin{align}
\mathbf x_k =  {\mathbf h_k} \theta_k + \boldsymbol \epsilon_k, ~ k = 1, 2, \ldots,
\label{eq: lin_sen}
\end{align}
where $\mathbf x_k \in \mathbb R^N$ is the vector of measurements acquired by $N$ sensors at time $k$, $\mathbf h_k \in \mathbb R^N$ is the vector of observation gains with known second-order statistics, $\mathbb {E}[{\mathbf{h}_k }] = \mathbf{h}$ and $\text{cov}({\mathbf{h}_k }) = \mathbf{\Sigma}_h$; 
$\boldsymbol \epsilon_k \in \mathbb R^N$ is the vector of measurement noises with zero mean and covariance  $\mathbb {E}[\boldsymbol \epsilon_{k_1}  \boldsymbol \epsilon_{k_2}^T ] = \mathbf{\Sigma}_{\epsilon} \mathcal I (k_1 - k_2)$. We also assume that $\{ \mathbf h_k \}$, $\{ \theta_k \}$ and $\{ \boldsymbol \epsilon_k \}$ are independent of each other.

After linear sensing, each sensor may pass its observation to
other sensors for collaboration, prior to transmission to the FC.
The sharing of measurements is accomplished via a coherent MAC. Collaboration among sensors is represented by a fixed  topology matrix $\mathbf A$ with binary entries, namely, $A_{mn} \in \{ 0, 1\}$ for $m \in [M]$ and $n \in [N]$,  where for ease of notation, $[M]$ denotes $\{1,2,\ldots,M \}$. Here $A_{mn}  = 1$ signifies that  there exists a communication link from  the $n$th sensor  to the $m$th sensor; otherwise,  $A_{mn}  = 0$. The neighborhood structure of the sensor network can be determined by the communication range of each sensor in the network.
Note
that $\mathbf A$ is essentially a truncated adjacency matrix with a self loop because each sensor
can collaborate with itself, namely, $\mathbf A_{mm} = 1$. 
We note that if  $\mathbf A_{mn} = 0$ for all
$m \neq n$, the resulting sensor collaboration strategy reduces to the amplify-and-forward transmission
strategy considered in \cite{xiao2008linear, jiang2014optimal}.

With a relabeling of   sensors, we assume that the first $M$ sensors (out of a total of $N$ sensor nodes) communicate with the FC. The condition $M \leq N$ is motivated by the need to reduce possible high transmission energy expense.
The prior knowledge about the network topology $\mathbf A$ is commonly determined based on the quality of observation and channel gains, and the   pairwise sensing distance \cite{kar2013linear}.

Given the network topology, the collaborative signal received at the $i$th sensor at time $k$ is  modeled as
\begin{align}
\begin{array}{l}
  \displaystyle   z_{k,i} = W_{k}(i,i) x_{k,i} + \left ( \sum_{j \in \mathcal N_i} W_{k}(i,j)  x_{k,j} + \kappa_{k,i} \right ),  \\
\mathcal N_i = \{ j ~| ~ A_{ij} = 1, ~ j \in [N] \ \text{and} \ j \neq i \},
\end{array}
\label{eq: col_sig}
\end{align}
where   $x_{k,i}$ is the $i$th entry of the observation vector $\mathbf x_k$, $\mathcal N_i$ is the set of immediate neighbors of the $i$th sensor, 
$W_{k}(i,i)$ is a power amplifying factor used for transmitting the measurement of the $i$th sensor to the FC, $ W_{k}(i,j)$  ($i \neq j$) is the power amplifying factor used for transmitting the measurement of the $j$th sensor to the $i$th sensor,  and $\{ \kappa_{k,i}, k \geq 0 \}$ denotes the communication  noise associated with a coherent MAC for inter-sensor collaboration, which is an independent, identically distributed (i.i.d) sequence, independent of other system parameters, with zero mean and variance $\sigma_{\kappa}^2$.


The sensor collaboration model \eqref{eq: col_sig} can also be  written in a compact form 
\begin{align}
\mathbf z_{k} =  \mathbf W_k  \mathbf x_k + \boldsymbol \kappa_{k} , ~\mathbf W_k \circ (\mathbf 1_M \mathbf 1_N^T - \mathbf A) = \mathbf 0,
\label{eq: col_sig_new}
\end{align}
where $\mathbf z_k = [z_{k,1}, \ldots, z_{k,M}]^T \in \mathrm R^M$ denotes the collaborative signals at time $k$,
$\mathbf W_k \in \mathbb{R}^{M \times N}$ is the collaboration matrix that contains collaboration weights (based on   the allocated energy) used to combine sensor measurements,  $W_{k}(i,j)$ is the $(i,j)$th entry of $\mathbf W_k$, $\boldsymbol \kappa_{k} = [ \kappa_{k,1} , \ldots, \kappa_{k,M}]^T  $ is the vector of receiver communication noises for sensor collaboration, $\circ$ denotes the Hadamard product, $\mathbf 1_N$ is the $N \times 1$ vector of all ones, and $\mathbf 0$ is the $M \times N$ matrix of all zeros. 
In what
follows, while referring to vectors of all ones and all zeros, their
dimensions will be omitted for simplicity but can be inferred
from the context.
In \eqref{eq: col_sig_new}, the second identity implies that $\mathbf W_k$ preserves the sparsity structure of the topology matrix $\mathbf A$.

After sensor collaboration, the collaborative signal $\mathbf z_k$ is transmitted through a coherent MAC so that the received signal $y_k$ at the FC is a coherent sum  
\begin{align}
y_k =  \mathbf g_k^T \mathbf z_k + \varsigma_k, ~  k = 1,2,\ldots,
\label{eq: channel}
\end{align} 
where $\mathbf g_k \in \mathbb R^M$ is the vector of  channel gains with known second-order statistics $\mathbb{E}[ {\mathbf{g}_k}] = \mathbf{g}$ and $\text{cov}( {\mathbf{g}_k}) = \mathbf{\Sigma}_g$,  
and $\{ \varsigma_k \}$ is an i.i.d noise sequence independent of $\{ \mathbf g_k \}$ and $\{ \mathbf z_k \}$ with zero mean and variance $\sigma_{\varsigma}^2$. 
We assume that the FC knows the second-order statistics
of the observation gain, channel gain, and additive noises.
Our goal is to design the optimal energy allocation scheme $\mathbf W_k$ such that the estimation distortion of tracking $\theta_k$ is optimized under
  real-time energy harvesting constraints.
Prior to formally stating our problem, we next introduce the transmission costs, energy-harvesting sensors, and the  estimation approach.

\begin{table}[]
\centering
\caption{List of System Parameters}
\label{table:list}
\begin{tabular}{| c | c | c | c |}
\hline
Parameter                              & Symbol                   & Mean        & Covariance                     \\ \hline
Parameter of interest                  & $\theta_k$               & 0           & $s_k$                        \\ \hline
State transition coefficient           & $\alpha$                 &   -          &     -                         \\ \hline
Parameter model noise & $\tau_k$                 &  0 & $\sigma_{\tau}^2$            \\ \hline
Observation gain vector                & $\mathbf h_k$            & $\mathbf h$ & $\mathbf{\Sigma}_h$          \\ \hline
Measurement noise vector               & $\boldsymbol \epsilon_k$ & $\mathbf 0$ & $\mathbf{\Sigma}_{\epsilon}$ \\ \hline
Collaboration noise at sensor $i$      & $\kappa_{k, i}$          & 0           & $\sigma_{\kappa}^2$          \\ \hline
Channel gain vector                    & $\mathbf g_k$            & $\mathbf g$ & $\mathbf{\Sigma}_g$          \\ \hline
Transmission noise                     & $\varsigma_k$            & 0           & $\sigma_{\varsigma}^2$     \\  \hline
Rayleigh fading channel parameter &$\xi$ & $\xi \sqrt{\frac{\pi}{2}}$  & $ \frac{4-\pi}{2} \xi^2$ \\  \hline
\end{tabular}
\end{table}

\subsection{Transmission cost}
In the entire estimation system, sensor energy is   consumed for  both inter-sensor communication  and sensor-to-FC communication.   
Based on \eqref{eq: col_sig},
the cost of  energy consumed at sensor $j \in [N]$  and time $k$   for  spatial collaboration is
\begin{align}
T^{(1)}_{k,j}(\mathbf W_k)   \Def \sum_{i \in \mathcal N_j}    (  W_{k}(i,j)  x_{k,j} )^2,  \label{eq: P_W1}
\end{align}
and its expectation over   stochastic parameters is given by 
\begin{align}
\mathbb E [  T^{(1)}_{k,j}(\mathbf W_k) ]  &   \overset{(a)}{=}    \left ( \mathbf e_j^T\mathbb E  [ \mathbf x_k \mathbf x_k^T ] \mathbf e_j  \right ) \sum_{i =1, i \neq j}^M W_{k}^2(i,j) \nonumber \\
 & =   \left ( \mathbf e_j^T\mathbb E  [ \mathbf x_k \mathbf x_k^T ] \mathbf e_j  \right ) 
 [ \mathbf e_j^T ( \mathbf W_k \circ \tilde{\mathbf I} )^T  ( \mathbf W_k \circ  \tilde{\mathbf I}  )  \mathbf e_j  ],  \label{eq: P_W12}
\end{align}
where $ \mathbb E  [ \mathbf x_k \mathbf x_k^T ] =\mathbf{\Sigma}_\epsilon + \sigma_\theta^2(\mathbf{h}\mathbf{h}^T + \mathbf{\Sigma}_h)$, the equality (a) holds since $W_{k}(i,j) = 0$ if $i \notin \mathcal N_j$ and $i \neq j$, $\circ$ denotes the Hadamard product, $\tilde{\mathbf I} \Def \mathbf 1 \mathbf 1^T -  [\mathbf I_M, \mathbf 0_{M \times (N - M)}]$, and 
$\mathbf e_j$ is   a basis vector
with $1$ at the $j$th coordinate and $0$s elsewhere.

Moreover from \eqref{eq: col_sig_new},  the cost of energy consumed at sensor $i \in [M]$ and time $k$ for sensor-to-FC communication becomes
\begin{align}
T^{(2)}_{k,i}(\mathbf W_k)   =  z_{k,i}^2 
\label{eq: P_W2}
\end{align}
and its expectation is given by
\begin{align}
\label{eq: P_W2E}
 \mathbb E [ T^{(2)}_{k,i}(\mathbf W_k)  ]=    
 \mathbf e_i^T  \mathbf W_k \mathbb E [ \mathbf x_k \mathbf x_k^T ] \mathbf W_k^T \mathbf e_i   + \sigma_{\kappa}^2.
\end{align}

The total transmission cost can now be represented as
\begin{align}
T_{k,i}(\mathbf W_k) = 
\left \{
\begin{array}{lr}
T^{(1)}_{k,i}(\mathbf W_k) + T^{(2)}_{k,i}(\mathbf W_k), & i \leq M,\\
T^{(1)}_{k,i}(\mathbf W_k),  & M < i \leq N,
\end{array}
\right.
\label{eq: P_W}
\end{align}
where we recall that
 only the first $M$ sensors are used to communicate with the FC,  and thus    $ T^{(2)}_{k,i}(\mathbf W_k)   = 0$ if $i > M$.

\subsection{Energy flow of energy harvesting sensors}
Let $H_{k,i}$ denote the harvested energy of the $i$th sensor at time $k$. Let $H_k = \{ H_{k,i}, i = 1, 2, \ldots, N \}$. Following \cite{sharma2010optimal, rajesh2014capacity}, we assume that $\{H_{k,i}, k \geq 0\}$ satisfy the strong law of large numbers (SLLN) with the limiting mean $\mathbb E [H_{k,i}] = \mu_i$: $\text{lim}\frac{1}{n}\sum_{k=1}^{n}H_{k,i} = \mu_i$ almost surely. An asymptotically stationary and ergodic process $\{ H_k \}$ satisfies this assumption. This energy  generation process is general enough to cover the assumptions usually made in the literature \cite{michelusi2012optimal, michelusi2013optimal, aprem2013transmit, kansal2007power, mao2013energy} and is quite realistic. In particular, this allows correlation in time of the energy harvested at a sensor as well as spatial correlation in the energy harvested at neighboring sensors. Moreover, we assume that the energy harvested
at the end of each time step, namely, $H_{k,i}$ is not  available until the beginning of time $k+1$.  The energy  storage  buffer of each sensor is considered to be large enough (here assumed with infinite capacity). 
And each sensor is supposed to know the energy  it has in its buffer at a given time. 
 We denote by $S_{k,i}$   the energy available in the  $i$th sensor at time $k$, which evolves as 
\begin{align}
{S_{ k+1,i} =    ( S_{k,i} - T_{k,i})^{+} + H_{k,i}, ~i\in [N],} 
\label{eq: dynamic_S}
\end{align}
where for notational simplicity,  we use $T_{k,i}$ to denote   the transmission cost given by \eqref{eq: P_W}, and $(x)^+ = \max \{ 0, x\}$.
 
\subsection{Problem Formulation}
In this setup, the goal is to solve the following problem: At each time $k$, find an energy allocation scheme $\mathbf W_k$ such that $\sum_{k = 1}^{\infty} \mathbb E [ (\theta_k - \hat {\theta}_k )^2 ]$ is minimized with the real-time energy constraints $T_{k,i} \leq S_{k,i}, \forall \, i = 1, 2, \ldots, N \,\, \text{and} \,\, \forall \, k \geq 1$.

In the above problem, $\hat {\theta}_k$ is the estimate of $\theta_k$. However, this problem may not remain finite for all choices of $\mathbf W_k$. Therefore, we consider the minimization of 
\begin{equation}
\label{eq: PD00}
\underset{N \to \infty}{\text{lim} \, \text{sup}} \, \frac{1}{N} \sum_{k = 1}^{N} \mathbb E [ (\theta_k - \hat {\theta}_k )^2 ].
\end{equation}
This is close to the standard average cost Markov decision problem (MDP) \cite{feinberg2012handbook} except that we have real-time constraints, which are not considered by the usual constrained MDP \cite{altman1999constrained}. Thus, we may consider a greedy approach: At each time $k$, find $\mathbf W_k$ that minimizes $\mathbb E [ (\theta_k - \hat {\theta}_k )^2 ]$ such that $T_{k,i} \leq S_{k,i}$ for each $i$. This will imply that we should take $T_{k,i} = S_{k,i}$ most of the time. But, this may not be a good strategy from a long term point of view because this would then mean $T_{k,i} = H_{k,i}$ most of the time and hence whenever $H_{k,i}$ is low, we may get high estimation error. Thus, a more sustainable strategy, as we will see later, will be $T_{k,i} \leq \text{min}(S_{k,i}, \mu_{k,i} - \eta)$, where $\eta$ is a small positive constant. Also, for computational reasons, because we need to perform the computations at each time $k$, we consider a linear estimator. Furthermore, ours is a distributed setup: the energy constraint $S_{k,i}$ will only be known to sensor $i$ at time $k$ and is randomly changing with time. Also, we would like the FC to compute $\mathbf W_k$ because it has the maximum computational power and is assumed to be connected to a regular power grid. However, sending the needed information $\mathbf h_k$, $\mathbf g_k$ and $\varsigma_k$ at each time $k$ will incur too much communication cost. Also, after computation, the FC will need to transmit $\mathbf W_k$ to all the sensors for them to decide on the collaboration strategy.

Taking the above considerations into account, in the following, we first formulate an offline optimization problem that can be solved using only the second-order statistics of $\mathbf h_k$, $\mathbf g_k$, $\mathbf H_k$, $\alpha$ and the variances of the receiver noises at different sensor nodes. This solution $\mathbf W^*$ will not necessarily satisfy the real-time energy constraint at the sensor nodes. The solution $\mathbf W^*$ is provided to all the sensor nodes and the FC before hand. Based on this solution, each sensor can obtain a real-time solution $\mathbf W_k$ using only its own energy information $S_{k,i}$ at time $k$. We will show that asymptotically the mean squared error (MSE) of the real-time solution matches that obtained when using $\mathbf W^*$. Furthermore, at the end, via simulations, we will also show that the performance of this scheme is close to that of an optimal linear estimator computed at the FC at a far more computational cost and using all the current information of $\mathbf h_k$ and $\mathbf g_k$.

\begin{remark}
The second-order statistics can be estimated using robust mean and variance estimators \cite{donoho1982breakdown}. Most commonly used robust estimators, e.g., M-estimators and S-estimators, to estimate the covariance matrix have been shown to have very low estimation error with a reasonal number of samples \cite{donoho1982breakdown}.
\end{remark}

\section{Offline Optimization Problem}
\label{sec: tracking}
Based on \eqref{eq: theta_k}\,--\,\eqref{eq: channel}, we consider the state space model at the FC,
\begin{align}
\begin{array}{l}
\theta_k = \alpha \theta_{k-1} + \tau_{k}, 
  \\
 y_k =  \mathbf g_k^T   \mathbf W_k {\mathbf h_k} \theta_k + v_k, 
\end{array}
\label{eq: state_space0}
\end{align}
where $v_k = \mathbf g_k^T \mathbf W_k \boldsymbol \epsilon_k + \mathbf g_k^T   \boldsymbol \kappa_{k} + \varsigma_k$. Here $v_k$ can be regarded as the measurement noise with statistics $\mathbb E [v_k] =  0$ and
\begin{align}
\label{eq: statistics_vk}
\cov(v_k) &= \tr (    \mathbf W_k \boldsymbol \Sigma_\epsilon \mathbf W_k^T \mathbb E [\mathbf g_k \mathbf g_k^T ] ) + \sigma_\kappa^2 \tr ( \mathbb E [\mathbf g_k  \mathbf g_k^T ] )  + \sigma_\varsigma^2 \nonumber \\
&= \tr (    \mathbf W_k \boldsymbol \Sigma_\epsilon \mathbf W_k^T \boldsymbol \Lambda_g ) + \sigma_\kappa^2 \tr ( \boldsymbol \Lambda_g )  + \sigma_\varsigma^2,
\end{align}
where $\boldsymbol \Lambda_g = \mathbb E [\mathbf g_k \mathbf g_k^T ] = \mathbf g \mathbf g^T + \boldsymbol \Sigma_g $ from \eqref{eq: channel}.

We will consider $\mathbf W_k = \mathbf W$ in this section and find $\mathbf W$ that minimizes \eqref{eq: PD00}. Since we are considering a linear estimator, we can formulate it as a Kalman filter. However, even with $\mathbf W_k = \mathbf W$, the filter for the state space model given in \eqref{eq: state_space0} is not a standard Kalman filter because of the \textit{multiplicative noise} $\mathbf g_k^T   \mathbf W {\mathbf h_k}$ in the measurement model. A recursive linear minimum mean squared error estimator (R-LMMSE) in the presence of multiplicative noise has been considered in \cite{rajsatsri71, Tug81}. Motivated by the results in \cite{rajsatsri71, Tug81} and since we have already assumed that $\left | \alpha \right | < 1$, the following Lemma shows that \eqref{eq: state_space0} satisfies the assumption in \cite{Tug81} and hence that the R-LMMSE to be derived is stable. In the following, we denote $u_k := \mathbf g_k^T   \mathbf W {\mathbf h_k}$.

\vspace{1.5mm}
\begin{lemma}
\label{lemma: uv}
Given the state space model in \eqref{eq: state_space0}, we have the following results: $u_k$ is uncorrelated with $v_k$. Moreover, $\{v_k\}$ is a zero mean, uncorrelated sequence.
\end{lemma}
\textbf{Proof}: See Appendix\ \ref{appendix: lemma1}. \hfill $\blacksquare$
\vspace{1.5mm}

Based on Lemma\,\ref{lemma: uv}, R-LMMSE \cite{Tug81} is employed to track the random parameter $\theta_k$, which is shown in Algorithm\ \ref{algo1}. 
\begin{algorithm}
Inputs: $\hat{\theta}_0 = 0, P_0$ assumed known
\begin{enumerate}
\item Prediction step:
\begin{align}
 \hat{\theta}_{k|k-1} = \alpha \hat \theta_{k-1}, \label{eq: predict_LMMSE}
\end{align}
with the predicted estimation error
\begin{align}
 P_{k|k-1} = \alpha^2 P_{k-1} + \sigma_\tau^2. \label{eq: predict2_LMMSE}
\end{align}
\item Updating step:
\begin{align}
 \hat{\theta}_{k} = \hat{\theta}_{k|k-1}  + d_k ( y_k  -   \mathbb E [ u_k] \hat{\theta}_{k|k-1} ), 
\end{align}
with the updated estimation error
\begin{align}
P_{k} = (1 - d_k \mathbb E [ u_k] ) P_{k|k-1} , \label{eq: error_Pk_LMMSE}
\end{align}
where $d_k$ is the estimator gain
\begin{align}
d_k = \frac{ P_{k|k-1}  \mathbb E [ u_k]   }{ ( \mathbb E [ u_k] )^2 P_{k|k-1}  + \cov(v_k) + \cov(u_k)  s_k },
\label{eq: qk_LMMSE}
\end{align}
and 
\begin{equation}
\label{eq: s_k}
s_k \Def \cov(\theta_k), s_k  = \alpha^2 s_{k-1}  + \sigma_{\tau}^2
\end{equation}
with known initial condition $  s_0$, $\mathbb E [ u_k]  = \mathbf g^T \mathbf W \mathbf h$ and
\begin{align}
\label{eq: statistics_u}
 \mathbb E [ u_k u_j] = \left \{
 \begin{array}{cl}
  \tr (\mathbf W \mathbf{\Lambda}_h \mathbf W^T \mathbf{\Lambda}_g)  & k = j,\\
  0 & k \neq j,
 \end{array}
 \right.
\end{align}
where 
  $\boldsymbol \Lambda_g$ has been defined in \eqref{eq: statistics_vk},  and
    $\boldsymbol \Lambda_h = \mathbb E [\mathbf h_k \mathbf h_k^T ] = \mathbf h \mathbf h^T + \boldsymbol \Sigma_h$. Also, $\mathrm{cov}(v_k)$ has been defined in \eqref{eq: statistics_vk} with $\mathbf W_k = \mathbf W$ and $\mathrm{cov}(u_k) =  \mathbb E [ u_k^2] - (\mathbb E [ u_k])^2$.
\end{enumerate}
	\caption{R-LMMSE with known second-order statistics of the system.}
	\label{algo1}
\end{algorithm}

\begin{remark}
It is clear from  \eqref{eq: predict_LMMSE}\,--\,\eqref{eq: error_Pk_LMMSE} that if the multiplicative noise $u_k$ is deterministic, namely, $\cov(u_k) = 0$, R-LMMSE reduces to the commonly used Kalman filtering algorithm.  
\end{remark}


From \cite{Tug81}, we obtain that the estimation error $\mathbb E[(\theta_k - \hat{\theta}_k)^2] = P_k \to P(\infty)$ and $\mathrm{cov}(\theta_k) = s_k \to s_{\infty}$. From \eqref{eq: s_k}, we obtain
\begin{align}
\label{eq: s_k2}
s_k = \alpha^{2k} s_0 + \sigma_{\tau}^2 \sum_{i=1}^k \alpha^{2i-2} = \alpha^{2k} s_0 + \sigma_{\tau}^2 \frac{1-\alpha^{2k}}{1 - \alpha^2}.
\end{align}

Using the fact that $\alpha^2 < 1$, we note that $s_{\infty} = \frac{\sigma_{\tau}^2 }{1 - \alpha^2}$. Moreover, $P_{\infty}$ satisfies the Riccati equation
\begin{align}
P_{\infty} = \frac{\alpha^2 P_{\infty} \bar d}{\bar c^2 P_{\infty}  + \bar d} + \sigma_\tau^2, \label{eq: Pinf}
\end{align}
where $\bar c = \mathbf g^T \mathbf W \mathbf h$, and 
$
\bar d = \tr (    \mathbf W \boldsymbol \Sigma_\epsilon \mathbf W^T \boldsymbol \Lambda_g) + \sigma_\kappa^2 \tr ( \boldsymbol \Lambda_g)  + \sigma_\varsigma^2 + \tr (    \mathbf W \boldsymbol \Lambda_h \mathbf W^T \mathbb \boldsymbol \Lambda_g )s_{\infty} - ( \mathbf g^T \mathbf W \mathbf h)^2 s_{\infty}
$.
Also, then \eqref{eq: PD00} equals $P(\infty)$. Thus, we would like to obtain the scheme $\mathbf W$ that minimizes $P(\infty)$ which at the same time also satisfies the energy constraint in the mean, i.e., $\mathbb E[T_{i}(\mathbf W)] \leq \mu_i - \eta$, for all $i$ and a small positive $\eta$, where $T_{i}(\cdot)$ denotes the total transmission cost as $k \to \infty$.

It is known from  \cite[Lemma\,1]{leong2011asymptotics}  that $P_{\infty}$ is a decreasing function of $\bar{c}^2/\bar{d}$. Thus, the optimal $\mathbf W$ can be obtained by solving the following optimization problem:
\begin{align}
\begin{array}{ll}
\displaystyle \maximize_{\mathbf W } &  f(\mathbf W)   \\
 \st & \mathbf W \circ (\mathbf 1 \mathbf 1^T - \mathbf A) = \mathbf 0, \\
& \mathbb E [ T_{i}(\mathbf W) ] \leq  \mu_{i} - \eta,   ~ i \in [N],
\end{array}
\label{eq: prob_off_1}
\end{align}
where $T_{i}(\mathbf W)$ is the transmission cost given by \eqref{eq: P_W}, the expectation is with respect to the stationary distribution of $\theta_k$, $\mu_{i} $ is the mean of the harvested energy at sensor $i$, $\eta > 0$ is a given small constant and 
\begin{align}
 f(\mathbf W) =  \frac{  ( \mathbf g^T   \mathbf W {\mathbf h})^2}{ \tr ( \mathbf W [\boldsymbol \Lambda_h s_{\infty} + \boldsymbol \Sigma_\epsilon] \mathbf W^T \boldsymbol \Lambda_g) + \sigma_\kappa^2 \tr ( \boldsymbol \Lambda_g )  + \sigma_\varsigma^2 }, \label{eq: fW}
\end{align}
where $s_{\infty} = \frac{\sigma_{\tau}^2 }{1 - \alpha^2}$.
In the offline optimization problem \eqref{eq: prob_off_1}, the last inequality constraint implies that the  energy cost cannot exceed the amount of harvested energy on an \textit{average}, where  $\eta$ can be regarded as the remaining average energy that is
deliberately kept in the energy storage buffer for further usage \cite{liutowards}. We observe that $\{ \theta_k \}$ is a geometrically ergodic Markov chain and the regeneration epochs of $\{ \theta_k \}$ and $\{ (T_k(\mathbf W), \theta_k) \}$ are the same. Therefore, $\{ (T_k(\mathbf W), \theta_k) \}$ is also a geometrically ergodic Markov chain. In particular, $T_k(\mathbf W)$ satisfies SLLN.

\begin{remark}
Note that the stated optimization problem in \eqref{eq: prob_off_1} is an offline sensor collaboration scheme. Therefore, it may not satisfy real-time energy constraints. 
\end{remark}

This optimal offline sensor collaboration scheme needs to be determined only once, and thus has significant computational merit. In the next section, we use the offline scheme to derive 
  an online solution that asymptotically
achieves the optimal $P_{\infty}$ subject to the real-time energy constraints. Some key equations are summarized in Table \ref{table:eq}.

\begin{table}[]
\centering
\caption{Key Equations}
\label{table:eq}
\begin{tabular}{|c|c|}
\hline
Equation Number & Problem Statement                           \\ \hline
(24)            & Offline optimization problem                \\ \hline
(36)            & Constructed online energy allocation policy \\ \hline
(38)            & Statistics based optimization problem       \\ \hline
(42)            & Full knowledge based online policy          \\ \hline
\end{tabular}
\end{table}

\section{Optimal Sensor Collaboration: from Offline to Online Solution}
\label{sec: solution}
In this section to solve the offline problem \eqref{eq: prob_off_1}, 
we begin by concatenating the
nonzero entries of a collaboration matrix into a collaboration
vector. 
  There are two benefits of using matrix vectorization:
a) the topology constraint   can be eliminated without loss
of performance, which results in a less complex problem; b) the
structure of nonconvexity is   easily characterized via such a
reformulation. We then derive the optimal solution of offline problem \eqref{eq: prob_off_1} with the aid of semidefinite programming. Given the optimal offline sensor collaboration scheme, we provide an online
scheme which is asymptotically equivalent to the optimal offline solution.

In the offline problem \eqref{eq: prob_off_1}, the only optimization variables are the nonzero entries of the collaboration matrix. To solve this problem, we begin by concatenating the nonzero entries of $\mathbf{W}$ (columnwise) into the collaboration vector
\begin{equation}
\mathbf{w} = [w_1, w_2, \ldots, w_L]^T \in \mathbb R^L, 
\label{eq: w_vec}
\end{equation}  
where $L$ is the total number of nonzero entries in the topology matrix $\mathbf A$. We also note that  there exists a one-to-one correspondence between the elements of $\mathbf w$ and $\mathbf W$, that is,
there exists a row index $m_l$ and a column index $n_l$
 such that $w_l = [\mathbf W]_{m_l n_l}$, where $[\mathbf X]_{ij} $ (or $X_{ij}$) denotes the $(i,j)$th entry of a matrix $\mathbf X$.

 Another important observation from the offline problem \eqref{eq: prob_off_1} is that both the objective function and the energy constraint   
 involve  quadratic matrix functions\footnote{A quadratic matrix function is a function $f : \mathbb R^{n\times r} \to  \mathbb R$ of the form
$f(\mathbf X) = \tr(\mathbf X^T \mathbf A \mathbf X) + 2 \tr(\mathbf B^T  \mathbf X) + c$ for certain coefficient matrices $
\mathbf A$, $\mathbf B$ and $\mathbf c$.}.
Inspired by this fact, we explore  
the   relationship   between $\mathbf W$ and $\mathbf w$ in  Proposition\,\ref{prop:Wtow}.

\vspace{1.5mm}
\begin{myprop}
\label{prop:Wtow}
Given a matrix $\mathbf W \in \mathbb R^{M\times N}$ and its columnwise vector $\mathbf w \in \mathbb R^L$ that only contains the nonzero elements of $\mathbf W$, the expressions of  $\mathbf b^T \mathbf W$ and $\tr(\mathbf C \mathbf W \mathbf D \mathbf W^T)$ can be equivalently expressed   as functions of $\mathbf w$, 
\begin{align}
&\mathbf b^T \mathbf W = \mathbf w^T \mathbf  B, \label{eq: lin12_w1} \\
& \tr(\mathbf C \mathbf W \mathbf D  \mathbf W^T) = \mathbf w^T \mathbf E \mathbf w, \label{eq: lin12_w2}
\end{align}
where   $\mathbf B \in \mathbb R^{L \times N}$ and $\mathbf E \in \mathbb R^{L \times L}$ are given by
\begin{align}
&B_{ln} = \left \{
\begin{array}{cc}
b_{m_l}, & n = n_l \\
0, & \text{otherwise}, 
\end{array}
\right.   \label{eq: A_w}\\
&  E_{k l} = [\mathbf C]_{m_k m_l} [\mathbf D]_{n_k n_l}, \label{eq: D_w}
\end{align}
for $n \in [N]$, $k \in [L]$ and $l \in [L]$, where the indices $m_l$ and $n_l$ satisfies
$w_l = W_{m_l n_l}$ for $l \in [L]$. 
\end{myprop}
\textbf{Proof:} See Appendix\ \ref{appendix: Prop2}. \hfill $\blacksquare$
\vspace{1.5mm}
 
 Based on Proposition \ref{prop:Wtow},  we can  simplify the offline problem \eqref{eq: prob_off_1} by: a) eliminating the network topology constraint, and b) expressing the objective function $f(\mathbf W)$ and the transmission cost $\mathbb E [T_i (\mathbf W)]$ in the forms that only involve quadratic vector functions. 
Consequently, the offline problem \eqref{eq: prob_off_1} becomes 
\begin{align}
\begin{array}{lll}
\vspace{1mm}
\displaystyle \maximize_{\mathbf w} & \displaystyle \frac{\mathbf w^T \boldsymbol{\Omega}_{\mathrm{N}} \mathbf w}{\mathbf w^T \boldsymbol{\Omega}_{\mathrm{D}} \mathbf w + \sigma_\kappa^2 \tr ( \boldsymbol \Lambda_g )  + \sigma_\varsigma^2},  & \\ 
\vspace{1mm}
\st &  \mathbf w^T \boldsymbol \Omega_{\mathrm T, i} \mathbf w  + \sigma_{\kappa}^2  \leq \mu_{i} - \eta,   & i \leq M,\\
& \mathbf w^T \boldsymbol \Omega_{\mathrm C, i}  \mathbf w \leq \mu_{i} - \eta, &  \hspace*{-0.3in} M < i \leq N,
\end{array}
\label{eq: prob1_off_equiv1}
\end{align}
where $ \boldsymbol{\Omega}_{\mathrm{N}} $, $\boldsymbol{\Omega}_{\mathrm{D}}$, $\boldsymbol \Omega_{\mathrm T, i}$ and $\boldsymbol \Omega_{\mathrm C, i}$ are all positive semidefinite matrices. We elaborate on the expressions of the aforementioned coefficient matrices in Appendix\ \ref{appendix: QuadV}. 

We remark that problem \eqref{eq: prob1_off_equiv1} is a nonconvex optimization problem as
it requires the maximization of a   ratio of convex quadratic functions. However, we will
show that semidefinite programming (SDP) can be used to find the
globally optimal solution of problem \eqref{eq: prob1_off_equiv1}.

By introducing a new variable $t$, together with a new constraint $1/t^2 = \mathbf w^T \boldsymbol{\Omega}_{\mathrm{D}} \mathbf w + \sigma_\kappa^2 \mathbf g^T  \mathbf g  + \sigma_\varsigma^2$, problem \eqref{eq: prob1_off_equiv1} can be rewritten as
\begin{align}
\begin{array}{ll}
\label{eq: prob1_off_equiv2}
\underset{\mathbf{w}, t}{\text{maximize}} \quad & t^2\mathbf w^T \boldsymbol{\Omega}_{\mathrm{N}} \mathbf w, 
\\ 
\vspace{1mm}
\st \quad & t^2\mathbf w^T \boldsymbol \Omega_{\mathrm T, i} \mathbf w + t^2\sigma_{\kappa}^2  \leq t^2 (\mu_{i} - \eta), ~ i \leq M,\\ 
\vspace{1mm}
& t^2\mathbf w^T\boldsymbol \Omega_{\mathrm C, i}\mathbf w \leq t^2(\mu_{i} - \eta),  ~~~ M < i \leq N, \\
\vspace{1mm}
& t^2\mathbf w^T \boldsymbol{\Omega}_{\mathrm{D}} \mathbf w + t^2\sigma_\kappa^2 \mathbf g^T  \mathbf g  + t^2\sigma_\varsigma^2 = 1,
\end{array}
\end{align}
where $t$ and $\mathbf w$ are optimization variables.

Upon defining 
$\bar{\mathbf{w}} = [
t\mathbf{w}^T , t
]^T \in \mathbb R^{L+1}$, problem \eqref{eq: prob1_off_equiv2} can be rewritten as
\begin{align}
\begin{array}{llr}
\label{eq: prob1_off_equiv3}
\underset{\bar{\mathbf{w}}}{\text{maximize}} \quad & \bar{\mathbf{w}}^T \mathbf Q_{0} \bar{\mathbf{w}},   & \\
\vspace{1mm}
\st \quad & \bar{\mathbf{w}}^T {\mathbf{Q}}_{i,1} \bar{\mathbf{w}} \leq 0,  & i \leq M, \\
\vspace{1mm}
& \bar{\mathbf{w}}^T {\mathbf{Q}}_{i,2}\bar{\mathbf{w}} \leq 0,  & M < i \leq N,  \\
\end{array}
\end{align}
\begin{align*}
\begin{array}{llr}
& \bar{\mathbf{w}}^T {\mathbf{Q}_{N+1}}\bar{\mathbf{w}} \leq 1, &
\end{array}
\end{align*}
where $\bar {\mathbf w}$ is the optimization variable, and
\begin{align*}
  & \mathbf{Q}_{0} = \begin{bmatrix}
\boldsymbol{\Omega}_{\mathrm{N}} & \mathbf{0} \\
\mathbf{0}^T & 0
\end{bmatrix} , ~~~~~~~\ \ \ \ \ \ 
\mathbf{Q}_{i,1} = \begin{bmatrix}
 \boldsymbol \Omega_{\mathrm T, i} & \mathbf{0} \\
\mathbf{0}^T & \sigma_{\kappa}^2 - (\mu_{i} - \eta) 
\end{bmatrix}, \\
& \mathbf{Q}_{i,2} = \begin{bmatrix}
\boldsymbol \Omega_{\mathrm C, i}  & \mathbf{0} \\
\mathbf{0}^T & - (\mu_{i} - \eta)
\end{bmatrix},~{\mathbf{Q}_{N+1}} = \begin{bmatrix}
\boldsymbol{\Omega}_{\mathrm{D}}  & \mathbf{0} \\
\mathbf{0}^T & \sigma_\kappa^2 \mathbf g^T  \mathbf g  + \sigma_\varsigma^2
\end{bmatrix}.
\end{align*} 
In problem \eqref{eq: prob1_off_equiv3}, 
 we  have replaced $\bar{\mathbf{w}}^T {\mathbf{Q}_{N+1}}\bar{\mathbf{w}} = 1$ with its inequality counterpart   without loss of performance since the  inequality constraint is satisfied with equality at the  solution while  maximizing a convex quadratic function. 
 We also note that the solution of  \eqref{eq: prob1_off_equiv1} can be obtained from the solution of problem \eqref{eq: prob1_off_equiv3}  via  
$\mathbf w = [\bar{\mathbf w}]_{1:L}/ [\bar{w}]_{L+1} $, where $[\bar{\mathbf w}]_{1:L}$ denotes the vector that consists of the first $L$ entries of $\bar{\mathbf w}$, and $[\bar{w}]_{L+1}$ is the $(L+1)$th entry of $\mathbf w$.

Problem \eqref{eq: prob1_off_equiv3} now contains a special nonconvex  structure: maximization of a  convex homogeneous quadratic function (i.e., no
linear term with respect to $\mathbf w$ is involved) subject to   homogeneous quadratic constraints. Although problem \eqref{eq: prob1_off_equiv3} is not convex, its globally optimal solution can be obtained via SDP 
\cite{jiang2014optimal,huamaizha10}. We provide the solution of problem \eqref{eq: prob1_off_equiv3} in Proposition\,\ref{prop: SDP_off}. In the following Proposition, $[\bar{\mathbf W}^*]_{L} = \mathbf{s}^*(\mathbf{s}^*)^T$ denotes the submatrix of $\bar{\mathbf W}^*$ formed by deleting its $(L + 1)$st row and column. It is rank $1$.

\vspace{1.5mm}
\begin{myprop}
\label{prop: SDP_off}
Let $\bar{\mathbf W}^*$ be the solution of the following SDP problem,
\begin{align}
\begin{array}{llr}
\label{eq: prob1_off_equiv5}
\underset{\bar{\mathbf W}}{\text{maximize}} \quad & \text{tr}({\mathbf{Q}_{0}}\bar{\mathbf W}), & \\
\vspace{1mm}
\st \quad & \text{tr}({\mathbf{Q}}_{i,1}\bar{\mathbf W}) \leq 0, & i \leq M,\\
\vspace{1mm}
& \text{tr}({\mathbf{Q}}_{i,2}\bar{\mathbf W}) \leq 0, &M < i \leq N, \\
\vspace{1mm}
& \text{tr}({\mathbf{Q}_{N+1}}\bar{\mathbf W}) \leq 1,~\bar{\mathbf W}  \succeq 0, & \vspace{-1mm}
\vspace{-2mm}
\end{array}
\end{align}
where $\bar{\mathbf W} \in \mathbb R^{(L+1) \times (L+1)}$ is the optimization variable. 
The solution of problem \eqref{eq: prob1_off_equiv1} is then given by
\begin{equation}
\mathbf{w}^* = \mathbf{s}^*/\sqrt{[{\bar W}^*]_{L+1, L+1}}, \label{eq: sol_wstar}
\end{equation}
where $[\bar{ W}^*]_{L+1, L+1}$ is  the $(L+1)$th diagonal element of $\bar{\mathbf W}^*$. 
\end{myprop}
\textbf{Proof:} See Appendix\ \ref{appendix: Prop3}. \hfill $\blacksquare$
\vspace{1.5mm}


The optimal offline sensor collaboration matrix, namely, the solution of the offline problem \eqref{eq: prob_off_1},
can now be found by using Proposition\,\ref{prop: SDP_off} and the sparsity pattern of the network topology matrix $\mathbf A$.
In what follows, we will
utilize the optimal offline sensor collaboration scheme to
design an online
sensor collaboration scheme $\mathbf W_k$ that satisfies the real-time energy
constraint $T_{k,i}(\mathbf W_k) \leq S_{k,i }$ for every time $k$ and every sensor $i$.
Here we recall that at time   $k$, each  sensor has access only to its stored energy $S_{k,n}$ given by \eqref{eq: dynamic_S}. 

We construct the following online energy allocation policy:
 \begin{align}
\begin{array}{ll}
\mathbf W_k  = \mathbf W^*,  & \text{if $T_{k,i}(\mathbf W^*) \leq S_{k,i}$, $\forall i \in [N]$}, \\
\mathbf W_k = \beta_{k}  \circ \mathbf W^*, &  \text{otherwise},
\end{array}
\label{eq: policy_given_off}
\end{align}
 where $\mathbf W^*$ denotes the optimal offline sensor collaboration matrix, $\circ$ denotes the Hadamard product, $\beta_k =  \min_{i} \{ \beta_{k,i}\}$, and 
$\beta_{k,i}  = \min\{1,\sqrt{S_{k,i}/{T_{k,i}(\mathbf W^*)}} \}$.
Once $\mathbf W_k$ is obtained from \eqref{eq: policy_given_off}, the transmission cost at time $k$ is then given by $T_{k,i} = T_{k,i}(\mathbf W_k)$. Subsequently,  the status of the energy storage buffer at the next time step is updated as $S_{k+1, i}$ according to \eqref{eq: dynamic_S}.

\begin{remark}
Note that the computation of  $\beta_k$ requires the global knowledge $\beta_{k,i}, i \in [N]$. 
However, we can use the max-consensus protocol \cite{iutzeler2012analysis} such that the computation  $\beta_k = \text{max}_i\{-\beta_{k,i}\}$ can be performed in a decentralized setting via inter-sensor message passing. 
\end{remark}

In Theorem\ \ref{prop: onlineoffline}, we show that the proposed online sensor collaboration policy is asymptotically consistent with the optimal offline solution in the mean-squared-error sense under the framework of R-LMMSE. To show this asymptotic consistency, we first present Lemma\ \ref{lemma: 2}.

\vspace{1.5mm}
\begin{lemma}
\label{lemma: 2}
As $k \to \infty$, the following results hold:
\begin{enumerate}
\item $\theta_k \overset{d}{\rightarrow} \theta_{\infty}, \mathbb E[\theta_k^{\gamma}] \to \mathbb E[\theta_{\infty}^{\gamma}]$ for $\gamma = 1, 2$, where $\overset{d}{\rightarrow}$ denotes convergence in distribution.
\item $\mathbb E[T_{k,i}(\mathbf W^*)] \to \mathbb E[T_i(\mathbf W^*)]$ for each $i \in [N]$.
\end{enumerate}
\end{lemma}
\textbf{Proof}: See Appendix\ \ref{appendix: lemma2}. \hfill $\blacksquare$
\vspace{1.5mm}

\begin{theorem}
\label{prop: onlineoffline}
Under the R-LMMSE framework \eqref{eq: predict_LMMSE}--\eqref{eq: error_Pk_LMMSE}, let 
 $\hat {\theta}_{k|k}$ and $\hat {\theta}^{\prime}_{k|k}$ denote the estimates 
when using the online sensor collaboration scheme \eqref{eq: policy_given_off} and the offline policy 
$\mathbf W^*$. Then
\begin{equation}
\label{eq: PD}
\underset{k \to \infty}{\text{lim}}\mathbb E [ (\hat {\theta}_{k|k} -\hat {\theta}^{\prime}_{k|k} )^2 ] = 0.
\end{equation}
 \end{theorem}
\textbf{Proof}: See Appendix\ \ref{appendix: Prop4}. \hfill $\blacksquare$
\vspace{1.5mm}

In practice, the system parameters $\alpha, \sigma^2_{\tau}, \sigma^2_{\varsigma}, \sigma^2_{\kappa}, \Sigma_{\epsilon}, \Lambda_g, \Lambda_h$ needed for obtaining the online scheme $\mathbf W_k$ in \eqref{eq: policy_given_off}, will need to be estimated. This will incur estimation error. The following proposition shows that the MSE $P_\infty$ of the offline scheme $\mathbf W^*$ is robust with respect to the estimation error. The same can be shown for $P_k$.

\vspace{1.5mm}
\begin{myprop}
\label{prop: robust}
The asymptotic MSE $P_\infty$ is a continuous function of the system statistics parameters 
$\alpha, \sigma^2_{\tau}, \sigma^2_{\varsigma}, \sigma^2_{\kappa}, \Sigma_{\epsilon}, \Lambda_g, \Lambda_h$.
\end{myprop}
\textbf{Proof}: See Appendix\ \ref{appendix: Proprobust}. \hfill $\blacksquare$
\vspace{1.5mm}

It is also of interest to get the rate of convergence of $\mathbb E[(\hat{\theta}_{k|k} - \theta_k)^2] \to P(\infty)$. We know from \cite{Tug81} that the rate of convergence for the offline scheme $\mathbf W^*$ is exponential. However, for the online filter, the rate will depend on the distributions of the channel gains and $\tau_k$. The following lemma provides information for such rates.

\vspace{1.5mm}
\begin{lemma}
\label{lemma: 3}
The rate of convergence of $\mathbb E[(\theta_{k} - \hat{\theta}_{k|k})^2]$ to $P_{\infty}$ depends on the rate $P(\frac{1}{k}\sum_{j=1}^{k} (H_{j,i} - T_{j,i}(\mathbf W^*)) \leq \varpi)$
and $P(\frac{T_{k,i}(\mathbf W^*)}{k} > \varpi)$ to go to $0$ as $k$ tends to infinity, where $\varpi$ is a small positive constant less than $\eta$.
\end{lemma}
\textbf{Proof}: See Appendix\ \ref{appendix: lemma3}. \hfill $\blacksquare$
\vspace{1.5mm}

Now we provide some conditions on the distributions of the  system parameters to obtain some rates of convergence of $\mathbb E[(\hat{\theta}_{k|k} - \theta_k)^2] \to P(\infty)$. We can show that $\theta_\infty$ has an exponential moment if $\tau_k$ has a positive density everywhere and has finite moment generating function in a neighborhood of $0$. Then if $h_{k,i}, \theta_\infty$ and $\epsilon_k$ have light tails, $T_{k,i}$ has light tail. If $h_{k,i}$ is finite valued, then  $h_{k,i}$ $\theta_\infty$ is light tailed. Otherwise, this condition can get violated even if $h_{k,i}$ is light tailed.  If $H_k$ is also strongly aperiodic, geometrically ergodic  Markov chain, we can, using concentration inequalities for Markov processes \cite{adamczak2015exponential} and the above Lemma, show that we obtain exponential rate of convergence of $\mathbb E[(\hat{\theta}_{k|k} - \theta_k)^2] \to P(\infty)$. Otherwise, one may only get a polynomial rate of convergence.

We emphasize that the proposed online policy has a significant
computational merit in that it is only required to solve the sensor
collaboration offline problem \eqref{eq: prob_off_1} once and then minimal extra computation \eqref{eq: policy_given_off} is required at each time. Theorem\,\ref{prop: onlineoffline} relates the
performance of the online strategy with that of the optimal offline strategy.

Note that we can also directly obtain a real-time optimal online sensor collaboration scheme by minimizing $P_k$, the estimation error when using R-LMMSE, subject to the expected transmission cost and scaling its solution using \eqref{eq: policy_given_off}. The corresponding optimization problem is given as
\begin{align}
\begin{array}{ll}
\displaystyle \minimize_{\mathbf W_k } &  P_k   \\
 \st & \mathbf W_k \circ (\mathbf 1 \mathbf 1^T - \mathbf A) = \mathbf 0, \\
& \mathbb E [  \ T_{k,i}(\mathbf W_k) ] \leq  \mu_{i} - \eta,   ~ i \in [N],
\end{array}
\label{eq: prob_online_1}
\end{align}
where \begin{align*}
P_k =  \frac{  ( \mathbf g^T   \mathbf W_k {\mathbf h})^2}{ \tr (    \mathbf W_k [\boldsymbol \Lambda_h s_k + \boldsymbol \Sigma_\epsilon] \mathbf W_k^T \boldsymbol \Lambda_g) + \sigma_\kappa^2 \tr ( \boldsymbol \Lambda_g )  + \sigma_\varsigma^2 }.
\end{align*}
By converting the collaboration matrix to the collaboration vector, 
statistics based problem \eqref{eq: prob_online_1} can be solved using the optimization approach proposed in Section\,\ref{sec: solution}. However, the computational complexity for this filter is very high since we need to solve the optimization problem at each time instant. We do not know the stability of this filter either. Thus, unlike the optimal online filter obtained above, this filter is not asymptotically optimal, but rather a greedy optimal scheme. We will see in Section\,\ref{sec: NR} that this filter does not perform better than the above proposed online filter.

To demonstrate the efficacy of the proposed online policy, in Section\,\ref{sec: NR}  we will also present a comparison between our proposed sensor collaboration scheme with
the online solution based on the full knowledge about the dynamical system, i.e., the realizations of observation and channels gains at each time.


\section{Comparison With CSI Based Sensor Collaboration}
\label{sec: comparison}
In this section, for the purpose of comparison, we assume 
that the channel state information, i.e., $\mathbf h_k$ and $\mathbf g_k$, of the state-space model \eqref{eq: state_space0} is available at the FC at time $k$.
In this ideal case,  R-LMMSE simplifies to the standard Kalman filter shown in Algorithm\ \ref{algo2}.
\begin{algorithm}
Inputs: $\hat{\theta}_0 = 0, P_0$ assumed known
\begin{enumerate}
\item Prediction: \eqref{eq: predict_LMMSE} and \eqref{eq: predict2_LMMSE}
\item Update:
\begin{align}
 \hat{\theta}_{k} = \hat{\theta}_{k|k-1}  + q_k ( y_k  -   \mathbf g_k^T   \mathbf W_k {\mathbf h_k}  \hat{\theta}_{k|k-1} ),
 \label{eq:k1}
\end{align}
with the updated estimation error
\begin{align}
P_{k} = (1 - q_k \mathbf g_k^T   \mathbf W_k {\mathbf h_k}) P_{k|k-1} , \label{eq:error_Pk}
\end{align}
where $q_k$ is the Kalman filter gain
\begin{align}
q_k = \frac{P_{k|k-1}  \mathbf g_k^T   \mathbf W_k {\mathbf h_k}  }{P_{k|k-1} ( \mathbf g_k^T   \mathbf W_k {\mathbf h_k})^2  + \mathbb E [v_k^2] },
\label{eq: q_k}
\end{align}
where $\mathbb E [v_k^2]  =  \mathbf g_k^T  \mathbf W_k \boldsymbol \Sigma_\epsilon \mathbf W_k^T\mathbf g_k + \sigma_\kappa^2 \mathbf g_k^T  \mathbf g_k  + \sigma_\varsigma^2$.
\end{enumerate}
	\caption{R-LMMSE with CSI.}
	\label{algo2}
\end{algorithm}

The sensor collaboration problem given the CSI about the estimation system is cast as
\begin{align}
\begin{array}{ll}
\vspace{1mm}
\displaystyle \maximize_{\mathbf W_k} & \displaystyle \frac{ (\mathbf g_k^T   \mathbf W_k {\mathbf h_k} )^2}{ \mathbf g_k^T  \mathbf W_k \boldsymbol \Sigma_\epsilon \mathbf W_k^T\mathbf g_k + \sigma_\kappa^2 \mathbf g_k^T  \mathbf g_k  + \sigma_\varsigma^2}  \\
\vspace{1mm}
\st &  \displaystyle \mathbb E[T_{k,i}(\mathbf W_k)] \leq \mu_{i} - \eta,   ~ i \in [N],\\
\vspace{1mm}
& \mathbf W_k \circ (\mathbf 1_M \mathbf 1_N^T - \mathbf A) = \mathbf 0.
\end{array}
\label{eq:prob0_equiv}
\end{align}
Similarly, by converting the collaboration matrix to the collaboration vector, the
CSI based problem \eqref{eq:prob0_equiv} can be solved using the optimization approach proposed in Section\,\ref{sec: solution}. However, unlike the Kalman filter in Section\,\ref{sec: solution} with $\mathbf W_k = \mathbf W$, the stability of the Kalman filter for the CSI based case has not been established in the literature.
Let $\mathbf W^*_k$ be the solution of the CSI based problem \eqref{eq:prob0_equiv} at time $k$. 
We then use \eqref{eq: policy_given_off} to construct the
 real-time sensor collaboration scheme $\mathbf W_{k}$.

This Kalman filter is a greedy scheme and not necessarily even asymptotically optimal. This is because in this setup we cannot use our approach used for the offline filter to optimize $P_\infty$ since stability of this Kalman filter is not known.  Another possible approach to obtain  an optimal filter could be via Markov decision theory. But this would be computationally infeasible even for a small number of sensors. Also, such CSI at the FC is not available in practice. Instead, we usually have access only to   the second-order statistics of the system parameters for parameter inference as in Section\,\ref{sec: solution}. 
In the next section, we will empirically show that in terms of estimation accuracy, the sensor collaboration policy proposed in Section\,\ref{sec: solution} performs better than the CSI based approach at low signal-to-noise ratios (SNRs), where the SNR refers to SNRs in terms of measurement noise and collaboration noise. Moreover, the former is computationally light, while the latter in addition to requiring much more information, also requires the solution of the CSI based problem \eqref{eq:prob0_equiv} at every time step, which thus introduces high computational complexity.


\section{Numerical Results}
\label{sec: NR}
In this section, we demonstrate the efficacy of our proposed   sensor collaboration schemes through numerical examples. The collaborative tracking system is shown in Fig. \ref{fig: Inf_prob_MComp}, where the vector of observation gains $\mathbf h_k$ and channel gains   $\mathbf g_k$ are chosen randomly  and independently from $\mathrm{Rayleigh}(\xi)$ distribution with $\xi = 1$. Therefore, we obtain   $\mathbf h = \mathbf g = \sqrt{\pi / 2} \mathbf{1}$ and $\mathbf{\Sigma}_h = \mathbf{\Sigma}_g = (4 - \pi) / 2 \mathbf{I}$. The observation noise $\mathbf{\epsilon}_k$ and the channel noise $ \varsigma_k$ are drawn from zero-mean Gaussian distributions with variances $\mathbf{\Sigma}_\epsilon = 0.5(\mathbf{I} + \mathbf{1}\mathbf{1}^T)$ and $\sigma_{\varsigma}^2 = 1$. Similarly, the vector of collaboration noises $\boldsymbol {\kappa}_k$ at time $k$ is assumed to be Gaussian with covariance $\mathbf{\Sigma}_{\kappa} = \sigma_{\kappa}^2 \mathbf{I}$, where unless specified otherwise, we set $\sigma_{\kappa}^2 = 0.5$. The parameter $\theta_k$ evolves as a Gauss-Markov process with Gaussian noise $\tau_k$ of variance $\sigma_{\tau}^2 = 0.5$. The state transition parameter is set as $\alpha = 0.8$. The initialized variance of $\theta_k$ denoted by $s_0$ is 100. We assume that the spatial placement and neighborhood structure of the sensor network is modeled as a random geometric graph, RGG($N, r$) \cite{freris2010fundamentals, kar2013linear}, where sensors are uniformly distributed over a unit square meter area and bidirectional communication links are possible only for pairwise distances at most $r$ meters, where $r$ refers to collaboration radius. Namely, $\mathbf A_{mn} = 1$ if the distance between the $m$th sensor and the $n$th sensor is less than $r$, otherwise it is $0$. Without loss of generality, we assume that $N = 10$ sensors are uniformly distributed over the region of interest and unless otherwise specified, we will assume $M = 10$ and $r = 0.6 m$.  At each sensor, the harvested energy $H_{k,i}$ is modeled as an Exponential random variable with mean $\mu_i = 10$ for $i \in [N]$. 
In the considered sensor collaboration problems  \eqref{eq: prob_off_1}  and   \eqref{eq:prob0_equiv}, we choose $\eta \in \{ 0.01, 1 \}$. 

For clarity, we   summarize the empirically studied sensor collaboration schemes as follows.
\begin{itemize}
\item The proposed offline sensor collaboration scheme  $\mathbf W^*$ is the solution of problem \eqref{eq: prob_off_1}.
\item The proposed online sensor collaboration scheme $\mathbf W_k$ is defined in \eqref{eq: policy_given_off}.
\item The online sensor collaboration scheme based on temporally-dependent statistics of system parameters $\mathbf W_k^{1}$ is the solution of problem \eqref{eq: prob_online_1} that satisfies \eqref{eq: policy_given_off}.
\item The CSI based  online greedy sensor collaboration scheme $\mathbf W_k^{2}$ is the solution of problem \eqref{eq:prob0_equiv} that satisfies \eqref{eq: policy_given_off}.
\end{itemize}

We remark that compared to our proposed sensor collaboration schemes $\mathbf W^*$ and 
$\mathbf W_k$, the schemes $\mathbf W_k^{1}$ and  $\mathbf W_k^{2}$ require the solution of the optimization problems at every time step. This naturally leads to high computational cost. The tracking performance for the above four schemes is measured through the empirical mean squared error (MSE), which is computed by simulating over 500 samples and taking average.

\begin{figure}[htb]
\centering
\includegraphics[width=8cm]{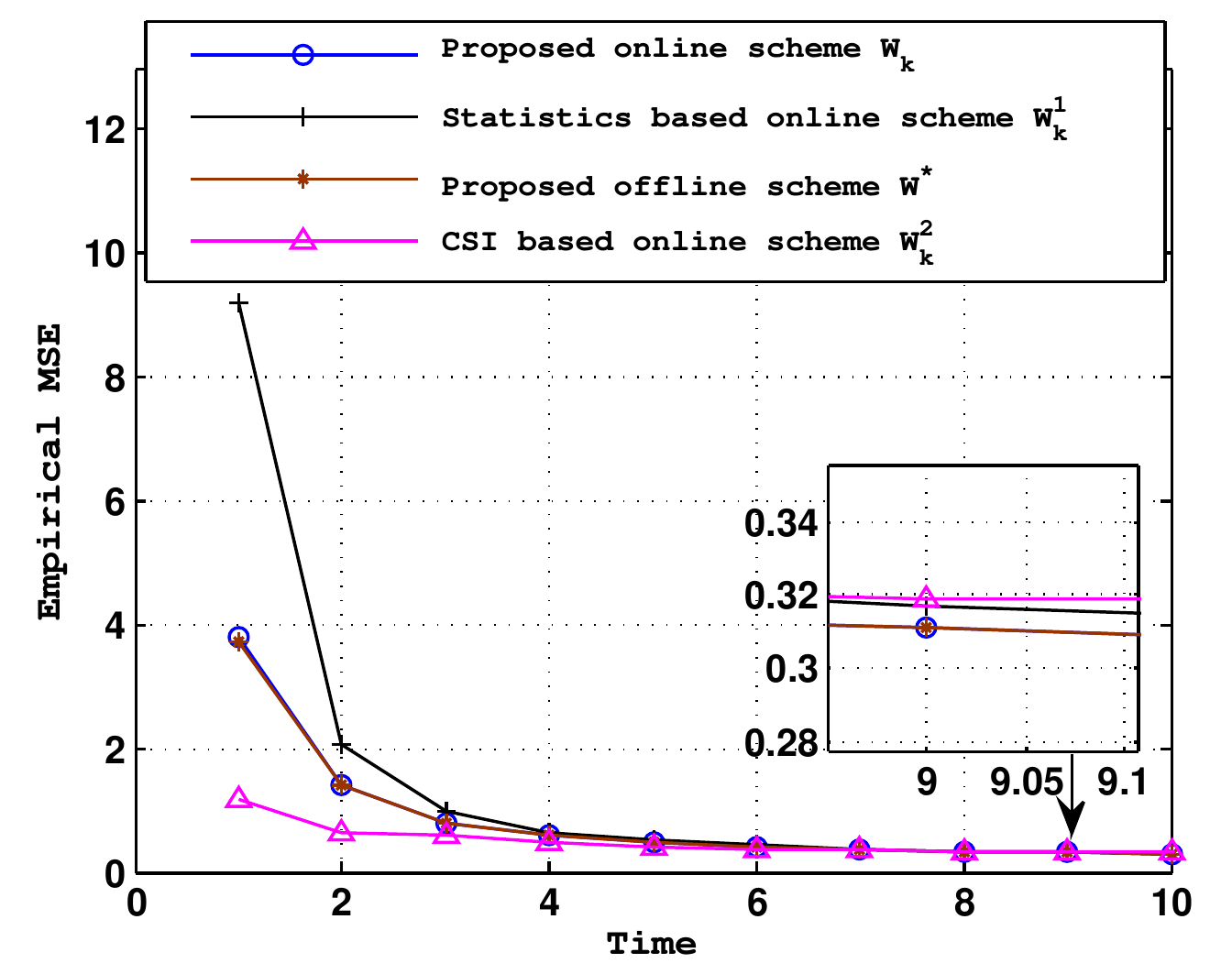} 
\caption{\footnotesize{Empirical MSE under $\eta = 0.01$.}}
\label{fig:E-mse}
\end{figure}

\begin{figure}[ht!]
\centering
\includegraphics[width=7.5cm]{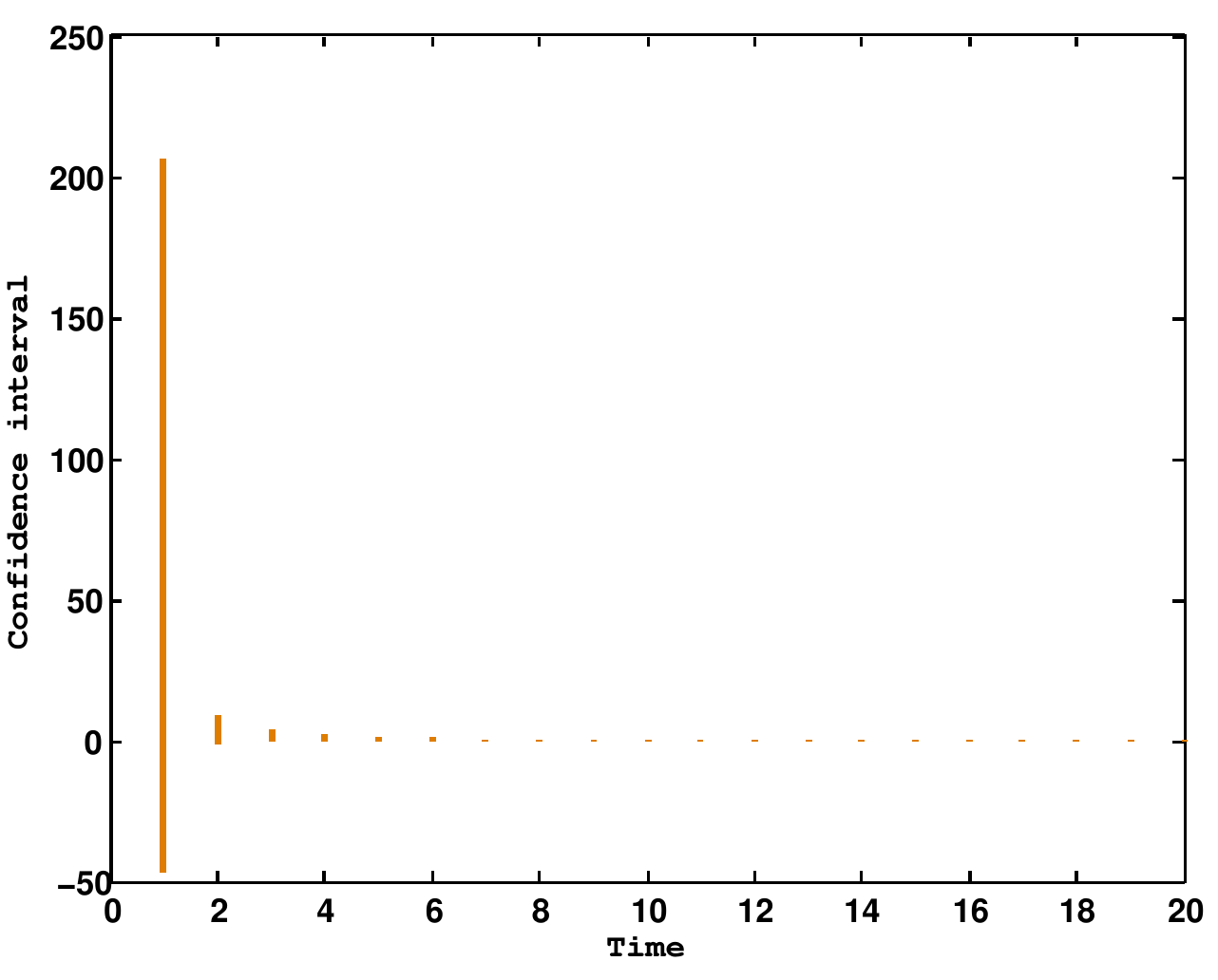} 
\caption{\footnotesize{$95\%$ confidence intervals of $\hat{\theta}_k$ for the proposed optimal online scheme under $\eta = 0.01$.}}
\label{fig:ci}
\end{figure}

\begin{figure}[ht!]
\centering
\includegraphics[width=8cm]{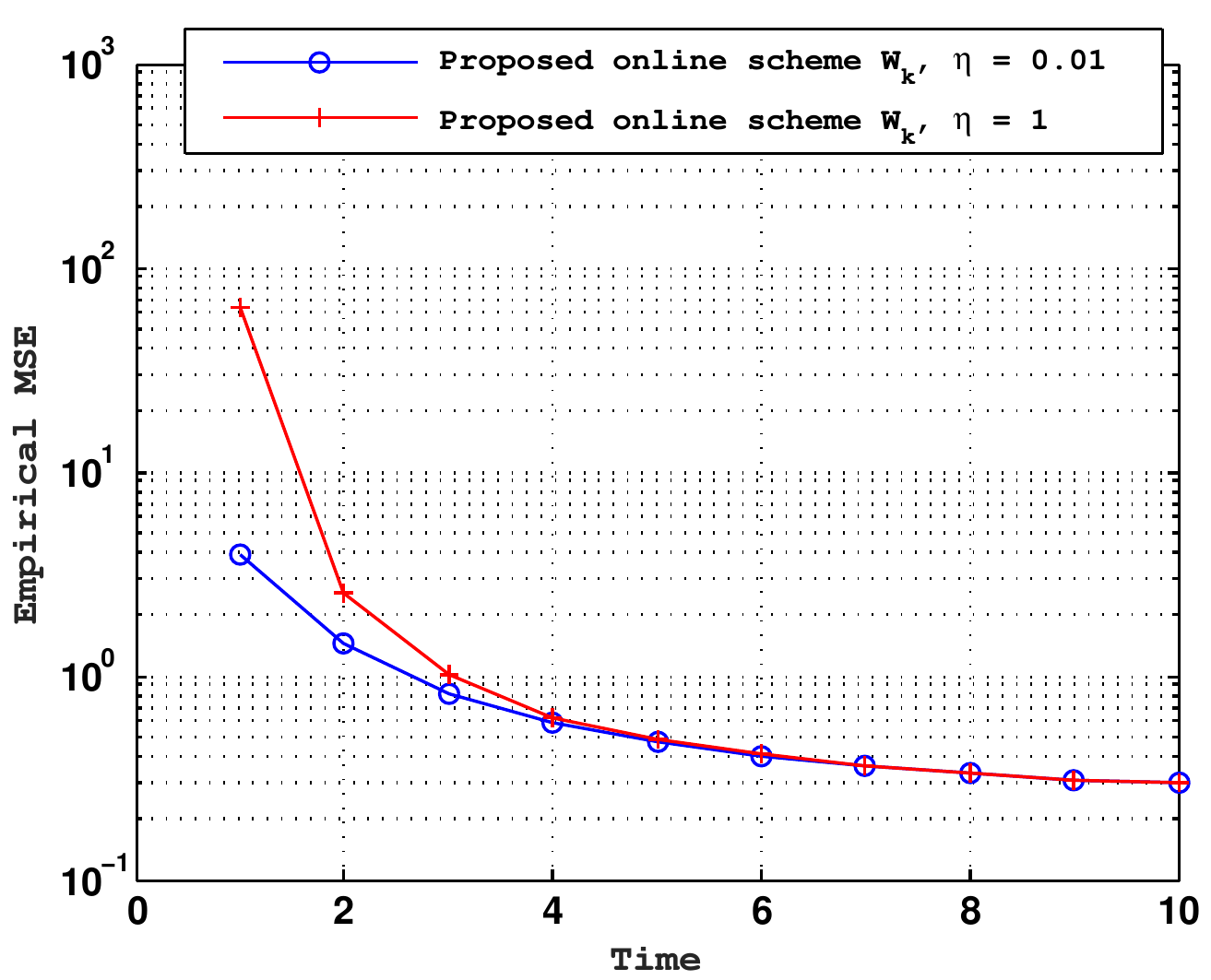} 
\caption{\footnotesize{Empirical MSE using the proposed online scheme $\mathbf W_k$ for   $\eta \in \{0.01,1 \}$.}}
\label{fig:E-mse2}
\end{figure}

\begin{figure}[ht!]
\centering
\includegraphics[width=8cm]{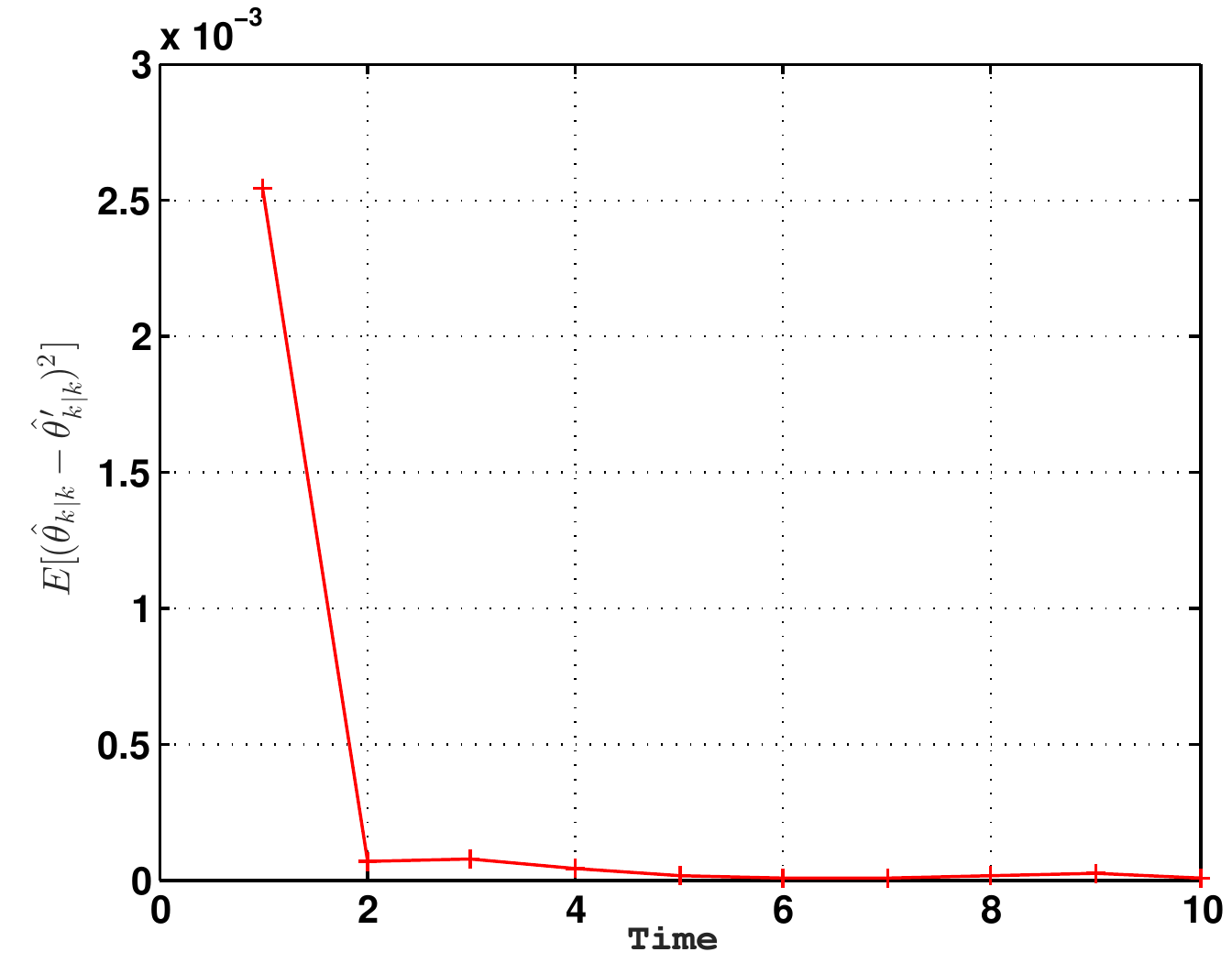} 
\caption{\footnotesize{Empirical verification of Theorem\ \ref{prop: onlineoffline}: Asymptotic consistency of the proposed online scheme and the optimal offline scheme under $\eta = 0.01$.}}
\label{fig:mmse}
\end{figure}

\begin{figure}[ht!]
\centering
\includegraphics[width=7.7cm]{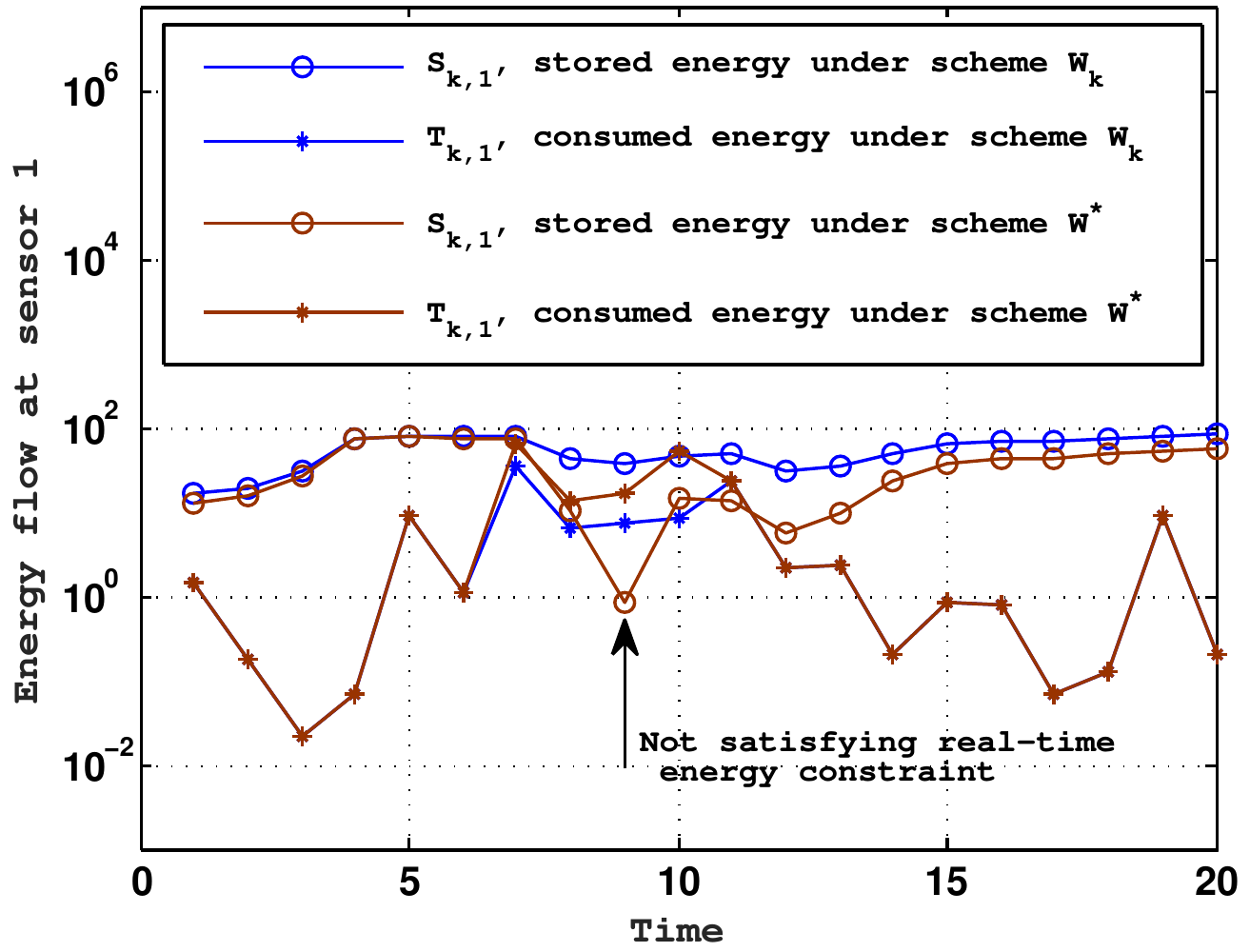} 
\caption{\footnotesize{Real-time energy flow.}}
\label{fig:energyflow}
\end{figure}

\begin{figure}[ht!]
\centering
\includegraphics[width=8.1cm]{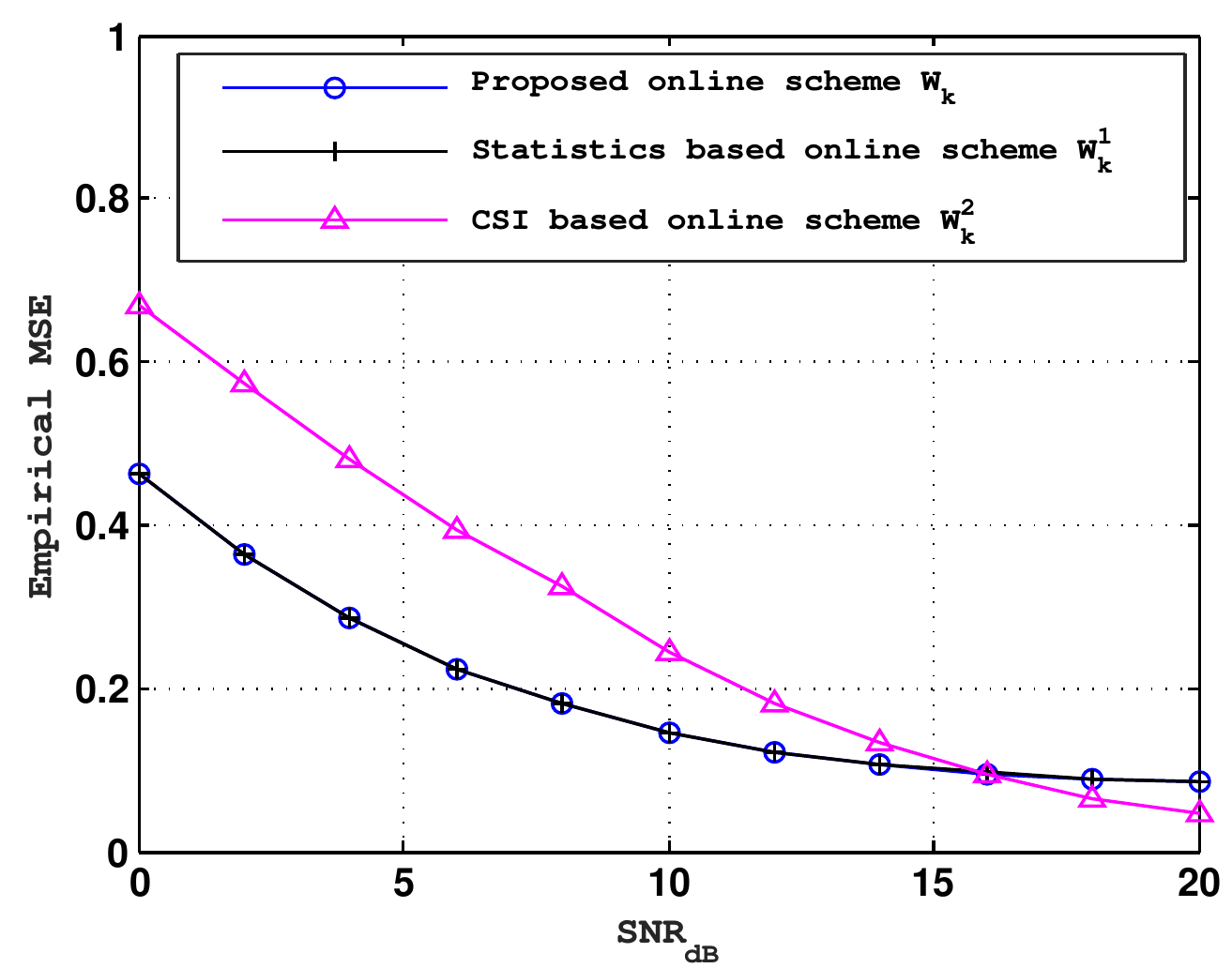} 
\caption{\footnotesize{Empirical MSE as a function of SNR in terms of measurement noise, $\eta = 0.01$.}}
\label{fig:SNR2}
\end{figure}

\begin{figure}[ht!]
\centering
\includegraphics[width=8cm]{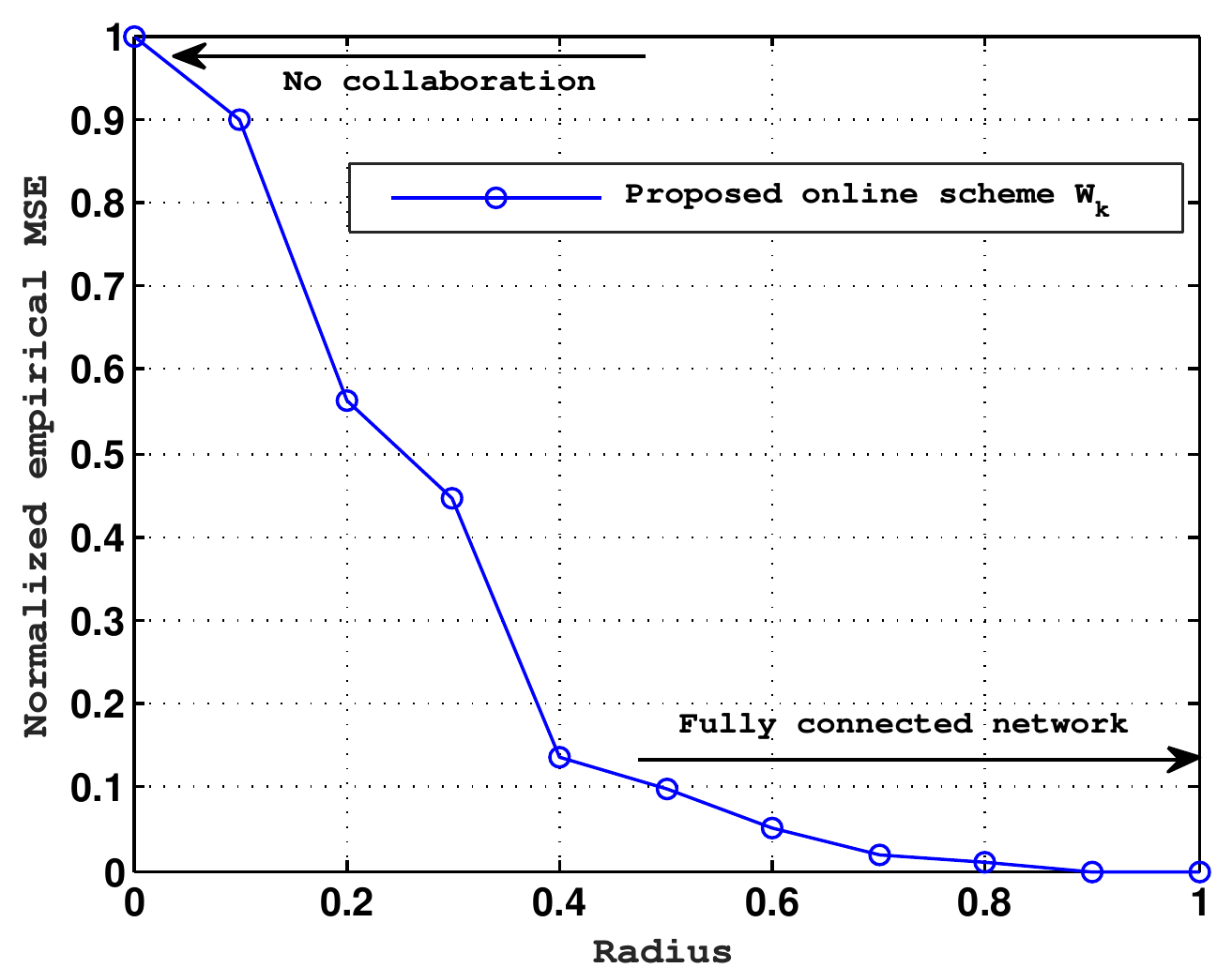} 
\caption{\footnotesize{Normalized empirical MSE as a function of collaboration radius.}}
\label{fig:radius}
\end{figure}

\begin{figure}[ht!]
\centering
\includegraphics[width=8cm]{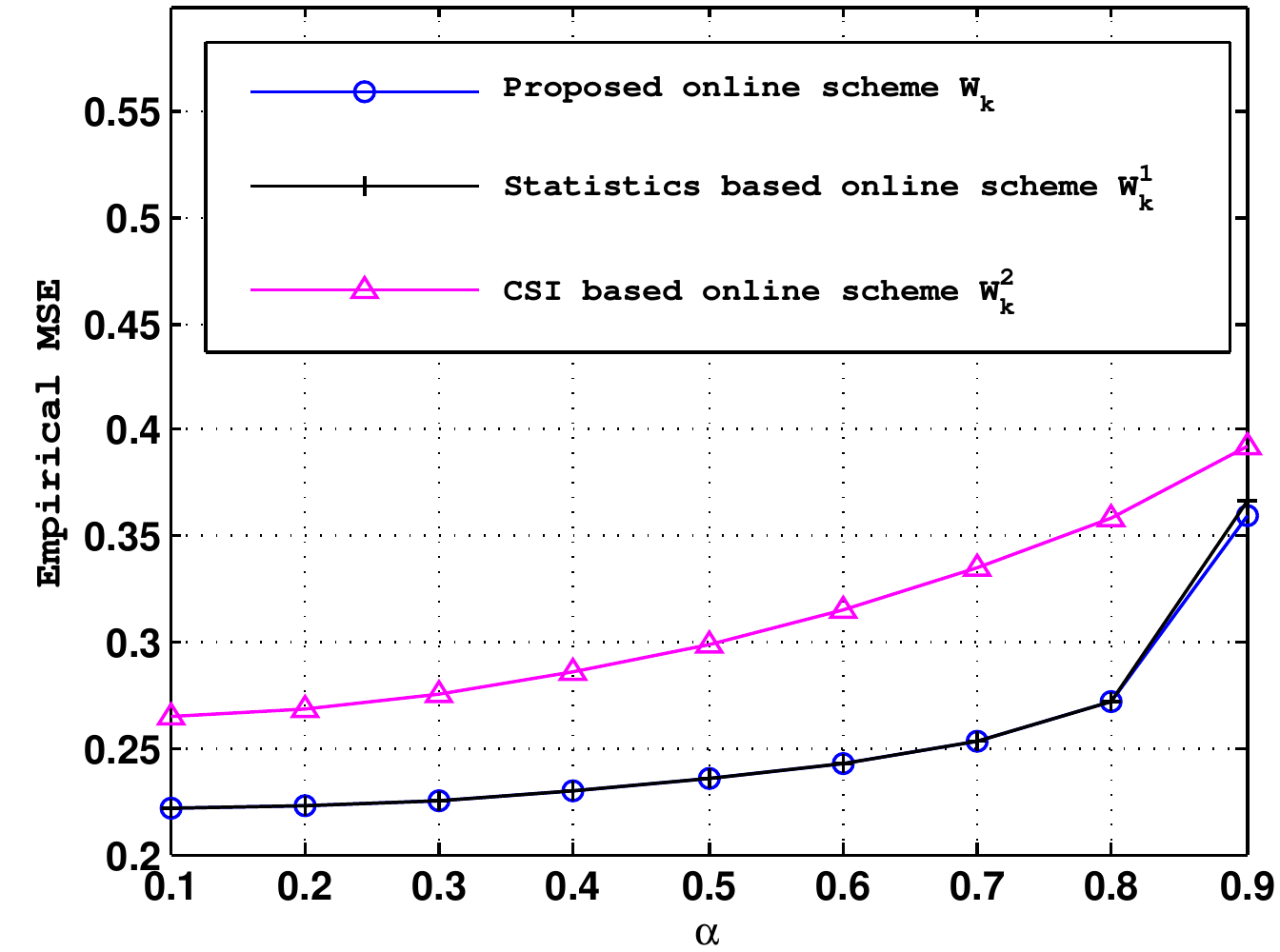} 
\caption{\footnotesize{Empirical MSE as a function of $\alpha$, $\eta = 0.01$ at low SNR in terms of measurement noise.}}
\label{fig:alpha}
\end{figure}

\begin{figure}[ht!]
\centering
\includegraphics[width=8cm]{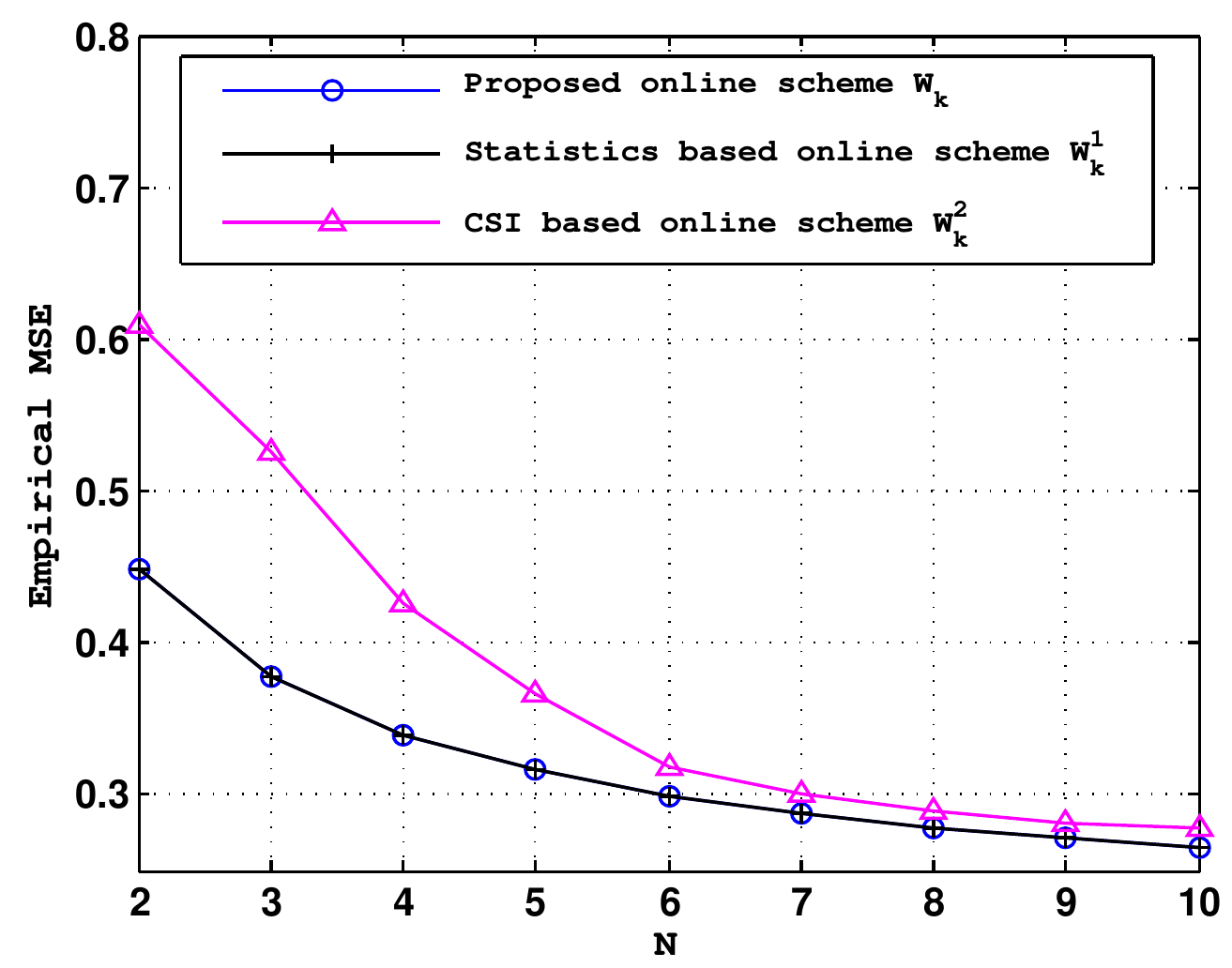} 
\caption{\footnotesize{Empirical MSE as a function of $N$, $\eta = 0.01$ at low SNR in terms of measurement noise.}}
\label{fig:N}
\end{figure}

\begin{figure}[ht!]
\centering
\includegraphics[width=8cm]{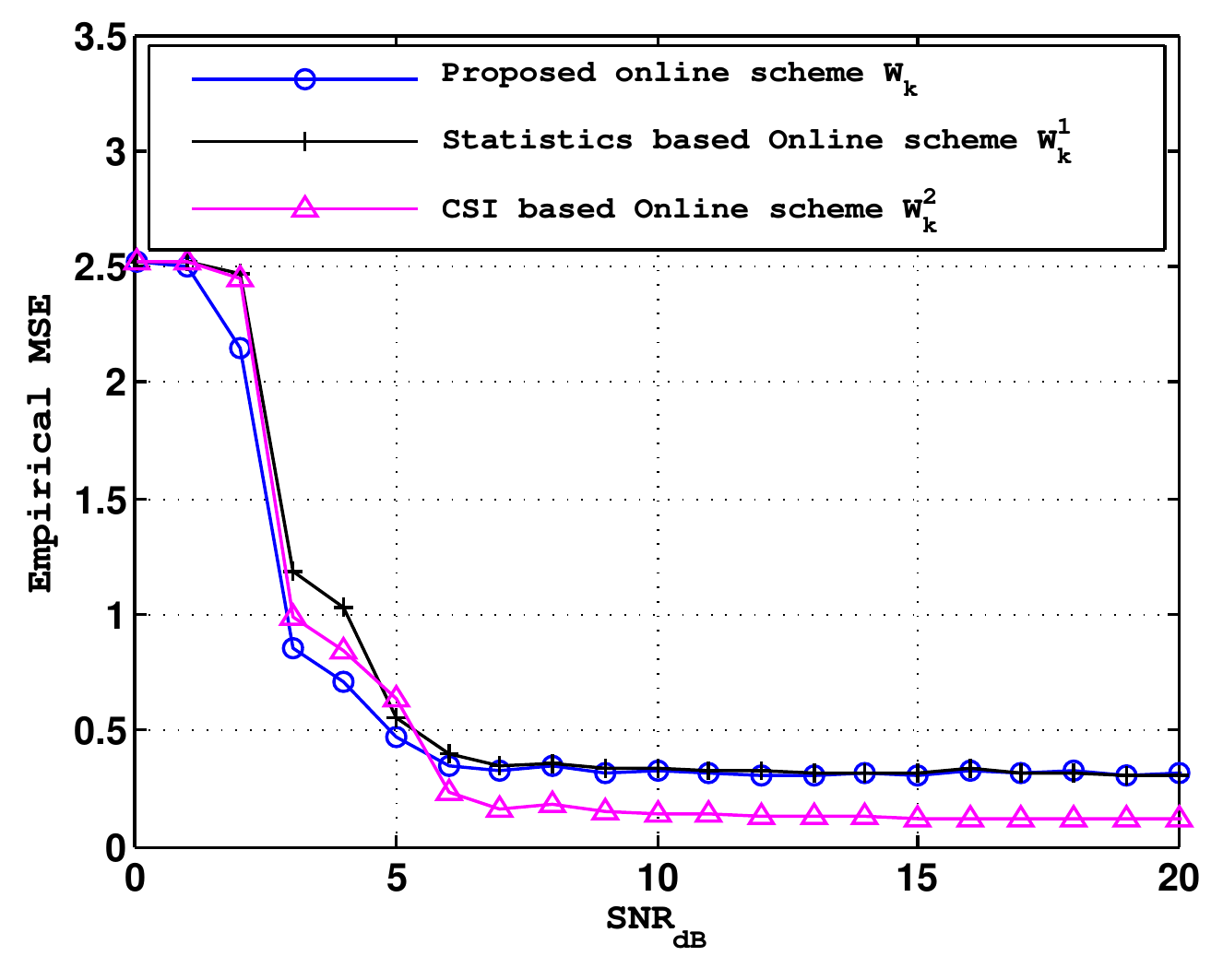} 
\caption{\footnotesize{Empirical MSE as a function of SNR in terms of collaboration noise, $\eta = 0.01$.}}
\label{fig:SNR1}
\end{figure}

\begin{figure}[ht!]
\centering
\includegraphics[width=8cm]{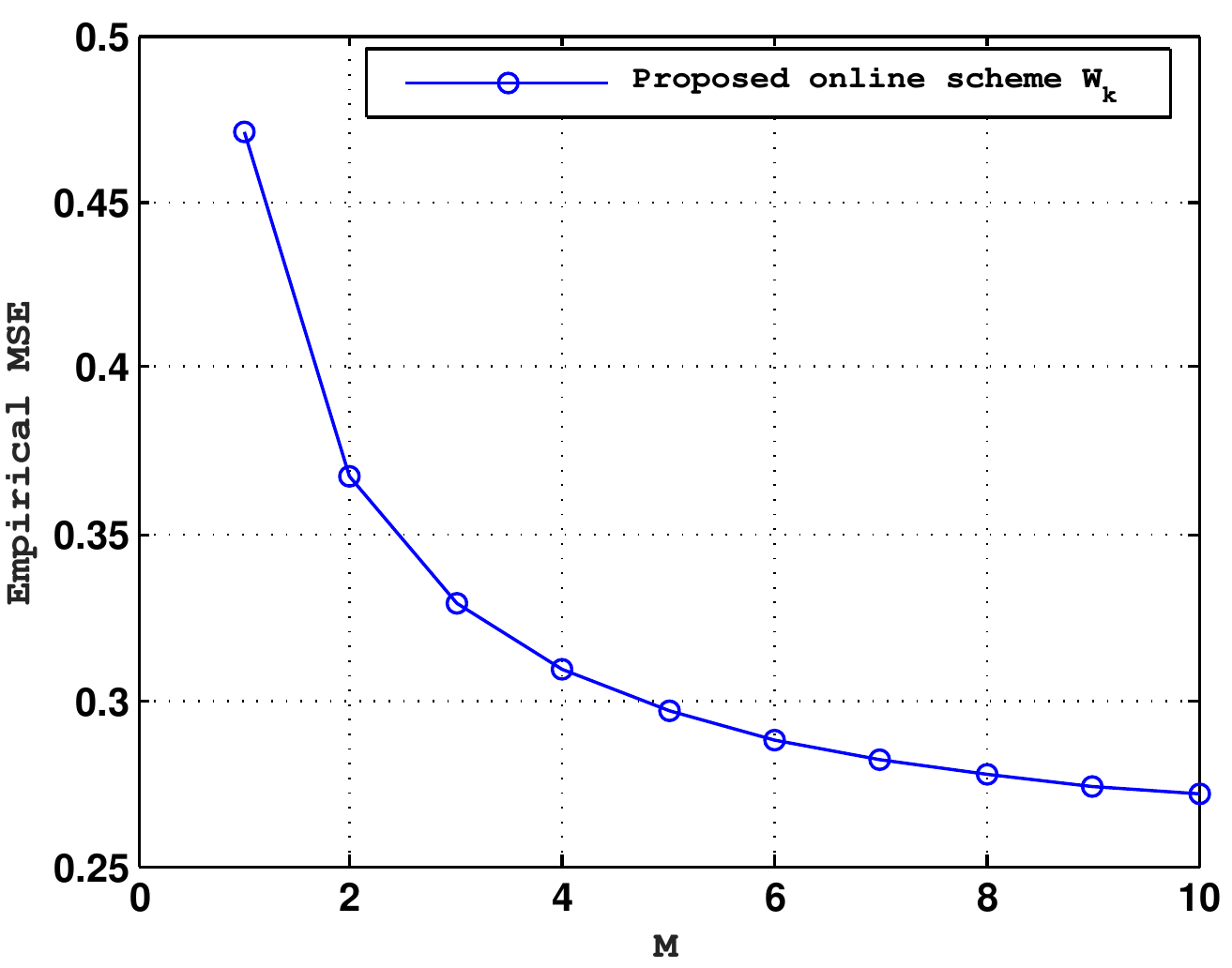} 
\caption{\footnotesize{Empirical MSE as a function of $M$ using the proposed online scheme $\mathbf W_k$, $\eta = 0.01$.}}
\label{fig:M}
\end{figure}


In Fig. \ref{fig:E-mse}, we present the empirical MSE when using the aforementioned sensor collaboration schemes as a function of time $k$ with $\eta = 0.01$. As we can see, the empirical MSE decreases as $k$ increases, and it converges to a steady-state value by $k = 10$. Note that the average variance of the parameter is decreasing as $k$ increases and eventually converges to a constant which implies that the average SNR is decreasing as $k$ increases; see \eqref{eq: s_k2}. As we can see, our proposed sensor collaboration schemes $\mathbf W^*$ and $\mathbf W_k$ yield MSE that is better than the greedy CSI based scheme $\mathbf W_k^{2}$ at low SNR. This indicates that our proposed scheme $\mathbf W_k$ is efficient not only in computation but also in estimation performance at low SNR. The empirical MSE against SNRs with respect to measurement noise and collaboration noise will be explored in the following. We also remark that the offline scheme $\mathbf W^*$ performs as well as the online schemes. But the former might not obey the real-time energy constraint.

In Fig. \ref{fig:ci}, we present the $95\%$ confidence intervals of $\hat{\theta}_k$ for the proposed optimal online scheme under $\eta = 0.01$ using two tailed t-test with a $1000$ number of samples. As we can see, our proposed online scheme performs very well as $k$ increases.

In Fig. \ref{fig:E-mse2}, we show the impact of the choice of $\eta$ on the tracking performance for the proposed online scheme. We recall that the parameter $\eta$ is the remaining average energy that is deliberately kept in the energy storage buffer. As we can see, the smaller the $\eta$ is, the better the tracking performance we obtain since more energy is available to be used in sensor collaboration. As a result, a larger $\eta$ leads to a less effective energy allocation scheme.

In Fig. \ref{fig:mmse}, we show the asymptotic performance of the proposed online scheme and the optimal offline scheme under $\eta = 0.01$, namely, when $k$ is large enough, the estimation performance of the proposed online scheme is the same as the offline scheme. As expected, the empirical MSE between $\mathbf W_k$ and $\mathbf W^*$ converges to $0$ as $k$ increases which is consistent with our theoretical result  in Theorem\ \ref{prop: onlineoffline}.

In Fig. \ref{fig:energyflow}, we present the real-time stored, and consumed energy at the first sensor for the proposed online schemes $\mathbf W_k$ and the proposed offline scheme $\mathbf W^*$ with $\eta = 0.01$. Here the stored energy and the consumed energy at sensor $1$ and time $k$ are denoted by $S_{k,1}$ and $T_{k,1}$, respectively. As we can see, the consumed energy $T_{k,1}$ is always less than or equal to the stored energy $S_{k,1}$ for the proposed efficient online scheme. Such dynamics of the energy flow is consistent with \eqref{eq: dynamic_S}. We have verified that the online schemes $\mathbf W_k^1$ and $\mathbf W_k^2$ also follow such energy dynamics. Moreover, we can see that the offline scheme violates real-time energy constraints.

In Fig. \ref{fig:SNR2}, we show the empirical MSE as a function of the average SNR in terms of measurement noise with $\eta = 0.01$ and $k = 20$. This $k$ is large enough to provide the converged estimation error. As we can see, the empirical MSE decreases when the average signal to measurement noise ratio increases. This is expected. By fixing the average signal to measurement noise ratio, we can see that the proposed online scheme $\mathbf W_k$ never performs worse than $\mathbf W_k^1$ which requires much more computational effort. Also, it performs better than the full information filter $\mathbf W_k^2$ for signal to measurement noise ratio less than $15$ dB.

In Fig. \ref{fig:radius}, we present the normalized empirical MSE as a function of the collaboration radius $r$ at time step $k = 20$ with $\eta = 0.01$. As we can see, the normalized empirical MSE decreases significantly as $r$ increases, since more collaboration links are established. Moreover, the normalized empirical MSE ceases to significantly decrease when $r > 0.7$. This indicates that a large part of the performance improvement is achieved only through partial collaboration.

In Fig. \ref{fig:alpha}, 
we show the empirical MSE as a function of $\alpha$ with $\eta = 0.01, k = 20$ for signal to measurement noise ratio less than $3$ dB. The empirical MSE increases when $\alpha$ increases. This is expected because the estimation error is a monotonic function of $\alpha$ that is the state transition coefficient of the parameter model; see \eqref{eq: error_Pk_LMMSE} - \eqref{eq: s_k}. By fixing $\alpha$, we can see that the proposed online scheme $\mathbf W_k$ performs close to or even better than the statistics based online scheme $\mathbf W_k^1$. Moreover, the proposed online scheme $\mathbf W_k$ performs better than the full knowledge based online scheme $\mathbf W_k^2$.

In Fig. \ref{fig:N}, we present the empirical MSE as a function of the number of sensors $N$ under $\eta = 0.01, k = 20$ for signal to measurement noise ratio less than $3$ dB. As we can see, the empirical MSE decreases when $N$ increases since more information is transmitted to the FC. Moreover, we can see the proposed online scheme $\mathbf W_k$ performs better than the full knowledge based online scheme $\mathbf W_k^2$.

In Fig. \ref{fig:SNR1}, we show the empirical MSE as a function of the average SNR in terms of collaboration noise with $\eta = 0.01$ and $k = 20$. As we can see, the empirical MSE decreases when the average signal to collaboration noise increases. This is expected. By fixing the average signal to collaboration noise, we can see that the proposed online scheme $\mathbf W_k$ never performs worse than statistics based online scheme $\mathbf W_k^1$ which requires much more computational effort. Also, it performs better than the full information filter $\mathbf W_k^2$ up to $5$ dB, which is the region of interest for energy harvesting sensor nodes \cite{sharma2010optimal}.

In Fig. \ref{fig:M}, we show the empirical MSE as a function of $M$, the number of sensors that communicate with the FC, for the proposed online scheme at $\eta = 0.01$ and $k = 20$. As we can see, the empirical MSE decreases when $M$ increases since more sensors are used for communication to the FC and, therefore the FC can acquire more information.


%
\section{Conclusion}
\label{sec: conclusion}
In this paper, we studied the problem of sensor collaboration for dynamic parameter tracking in energy harvesting sensor networks. We incorporated the effect of sensor collaboration noise while designing the online collaboration scheme. Based   on the second-order statistics of the system parameters, we   developed an online sensor collaboration policy that is built on the solution of the offline sensor collaboration problem. Although the latter is a complex non-convex optimization problem, we showed that it can be solved exactly via  semidefinite programming. We also proved that the proposed online policy is asymptotically consistent with the optimal offline solution over an infinite time horizon. We have also shown via simulations  that at low SNRs, in the region of interest in sensor networks, our online scheme actually outperforms a greedy optimal online policy which has full information at the fusion center about the channel gains and energy at the sensors at any given time. Numerical results have illustrated the efficiency of our approach and the impact of energy harvesting and collaboration noise on the  performance of collaborative estimation. 

In future work, it would be desirable to consider a more practical sensor collaboration model by
taking into account the impact of pairwise distances between sensors on their channel gain. Also, it is interesting to take into account the energy consumption and delay due to the estimation of second-order statistics. Moreover, the scalar random parameter can be generalized to a vector of random parameters. Also, one can study quantization-based inter-sensor collaboration. Lastly, the design of   network topologies could be taken into account under the current framework.


\appendices

\section{Proof of Lemma \ \ref{lemma: uv}}
\label{appendix: lemma1}
Since $u_k = \mathbf g_k^T   \mathbf W {\mathbf h_k}$ and $v_k = \mathbf g_k^T \mathbf W \boldsymbol \epsilon_k + \mathbf g_k^T   \boldsymbol \kappa_{k} + \varsigma_k$, we have
\begin{align}
\label{eq: uv}
\mathbb E[u_k v_k] &= \mathbb E[\mathbf g_k^T   \mathbf W \mathbf h_k (\mathbf g_k^T \mathbf W \boldsymbol \epsilon_k + \mathbf g_k^T   \boldsymbol \kappa_{k} + \varsigma_k)] \\ \nonumber
&= \mathbb E[\mathbf g_k^T   \mathbf W \mathbf h_k \mathbf g_k^T \mathbf W \boldsymbol \epsilon_k] + \mathbb E[\mathbf g_k^T   \mathbf W \mathbf h_k \mathbf g_k^T   \boldsymbol \kappa_{k}] \\ \nonumber
&+ \mathbb E[\mathbf g_k^T   \mathbf W \mathbf h_k \varsigma_k]. \nonumber
\end{align}
The first term on the right hand side (RHS) of \eqref{eq: uv} yields
\begin{equation*}
\mathbb E[\mathbf g_k^T   \mathbf W \mathbf h_k \mathbf g_k^T \mathbf W \boldsymbol \epsilon_k] = \mathbb E[\mathbf g_k^T   \mathbf W \mathbf h_k \mathbf g_k^T \mathbf W] \times \mathbb E[\boldsymbol \epsilon_k] = 0.
\end{equation*}

Applying similar analysis for other terms on the RHS of \eqref{eq: uv}, we can conclude that $u_k$ is uncorrelated with $v_k$. Likewise, we can show that $\{ v_k \}$ is an uncorrelated sequence. \hfill $\blacksquare$

\section{Quadratic Vector Functions}
\label{appendix: QuadV}
Based on Proposition \ref{prop:Wtow}, the offline problem \eqref{eq: prob_off_1} can be simplified such that it only contains quadratic vector functions with respect to $\mathbf w$. Using the identities \eqref{eq: lin12_w1} and \eqref{eq: lin12_w2}, we can rewrite the quadratic matrix functions in $f(\mathbf W)$ as
\begin{align}
&(\mathbf g^T\mathbf W {\mathbf h})^2 = \mathbf w^T \boldsymbol{\Omega}_{\mathrm{N}} \mathbf w, \label{eq: OmegaN} \\
&\tr (\boldsymbol \Lambda_g \mathbf W [\boldsymbol \Lambda_h s_{\infty} + \boldsymbol \Sigma_\epsilon] \mathbf W^T)  = \mathbf w^T \boldsymbol{\Omega}_{\mathrm{D}} \mathbf w, \label{eq: OmegaD}
\end{align}
where $\boldsymbol{\Omega}_{\mathrm{N}} =  \mathbf G\mathbf h \mathbf h^T \mathbf G^T$, $\mathbf G$ is defined as in \eqref{eq: A_w} that satisfies $\mathbf g^T \mathbf W = \mathbf w^T \mathbf G$, the $(k,l)$th entry of $\boldsymbol{\Omega}_{\mathrm{D}}$ is defined as in \eqref{eq: D_w}, namely, $[\boldsymbol{\Omega}_{\mathrm{D}}]_{kl} =  [\boldsymbol \Lambda_g]_{m_k m_l} [\boldsymbol \Lambda_h s_{\infty} + \boldsymbol \Sigma_\epsilon]_{n_k n_l}$.

Recall the expected transmission cost for sensor-to-FC in \eqref{eq: P_W2E}, 
\begin{align}
\label{eq: OmegaT2}
 \mathbb E [ T^{(2)}_{i}(\mathbf W)  ] &=    
 \mathbf e_i^T  \mathbf W [\boldsymbol \Lambda_h s_{\infty} + \boldsymbol \Sigma_\epsilon] \mathbf W^T \mathbf e_i   + \sigma_{\kappa}^2 \\ \nonumber
&=\text{tr}(\mathbf e_i \mathbf e_i^T\mathbf W [\boldsymbol \Lambda_h s_{\infty} + \boldsymbol \Sigma_\epsilon] \mathbf W^T) \\ \nonumber
&= \mathbf w^T \boldsymbol{\Omega}_{i}^{\mathrm{T}(2)}\mathbf w, 
\end{align}
where $\boldsymbol{\Omega}_{i}^{\mathrm{T}(2)}$ is also obtained according to \eqref{eq: D_w}, $[\boldsymbol{\Omega}_{i}^{\mathrm{T}(2)}]_{k l} = [\mathbf e_i \mathbf e_i^T]_{m_k m_l}[\boldsymbol \Lambda_h s_{\infty} + \boldsymbol \Sigma_\epsilon]_{n_k n_l}$.

According to \eqref{eq: P_W12}, the expected transmission cost for inter-sensor communication is a function of  $\mathbf W \circ  \tilde{\mathbf I}$, where $\tilde{\mathbf I} = \mathbf 1 \mathbf 1^T - [\mathbf I_M, \mathbf 0_{M \times (N - M)}]$. We further define $\tilde{\mathbf W} = \mathbf W \circ (\mathbf 1 \mathbf 1^T - [\mathbf I_M, \mathbf 0_{M \times (N - M)}])$, and one can view the matrix $\tilde{\mathbf W}$ as the collaboration matrix $\mathbf W$ with zero diagonals. Therefore, 
\begin{equation}
\label{eq: E-s-s-C}
\mathbb E[T^{(1)}_{i}(\mathbf W)] = \text{tr}(\mathbf e_i \mathbf e_i^T [\boldsymbol \Lambda_h s_{\infty} + \boldsymbol\Sigma_{\epsilon}]) \text{tr}(\tilde{\mathbf W} \mathbf e_i \mathbf e_i^T\tilde{\mathbf W}^T).
\end{equation}

Similar to $\mathbf w$, let $\tilde{\mathbf w}$ be the vector that consists of the nonzero entries (columnwise) of $\tilde{\mathbf W}$, namely, $\tilde{w}_l = W_{m_l n_l}$ for certain indices $m_l \in [M]$, $n_l \in [N]$ and $m_l \neq n_l$. Clearly, $\tilde{\mathbf w}$ is a vector of $(L-M)$ dimension, and it can be readily expressed as a linear function of $\mathbf w$, namely,
$\tilde{\mathbf w} = \mathbf F \mathbf w$ with $\mathbf F \in \mathbb{R}^{(L-M) \times L}$, given by
\begin{align}
F_{ij} = \left \{
\begin{array}{cc}
1, & \mathcal{K} (i) = j, \quad j \in [L], i \in [L - M], \\
0, & \text{otherwise}, 
\end{array}
\right.   
\label{eq: F}
\end{align}
where $\mathcal{K} = \{ l | w_l = W_{m_l n_l}, m_l \neq n_l \}$. 

We elaborate this linear transformation through the following example.
\begin{align}
\mathbf A _1= \begin{bmatrix}
 1& 0 &1  \\ 
0 & 1 & 1\\ 
 1& 0 & 1
\end{bmatrix}, \mathbf w = \begin{bmatrix}
W_{11}\\ 
W_{31}\\ 
W_{22}\\ 
W_{13}\\ 
W_{23}\\ 
W_{33}
\end{bmatrix}, \tilde{\mathbf w} = \begin{bmatrix}
W_{31}\\ 
W_{13}\\ 
W_{23}
\end{bmatrix}
\end{align}
Given $\mathbf A _1$, we can obtain that $\mathcal{K} = \{2, 4, 5\}$. Therefore, $F_{12} = 1, F_{24} = 1$ and $F_{35} = 1$, and otherwise $F_{ij} = 0$. We have 
\begin{equation*}
\mathbf F _1= \begin{bmatrix}
 0& 1 &0&0&0&0  \\ 
0 & 0 & 0&1&0&0\\ 
 0& 0 & 0&0&1&0
\end{bmatrix}.
\end{equation*}

By concatenating $\tilde{\mathbf W}$ to $\tilde{\mathbf w}$ and using the identity \eqref{eq: lin12_w2} as well as the linear transformation $\tilde{\mathbf w} = \mathbf{F}\mathbf w$, we can rewrite the quadratic matrix function in \eqref{eq: E-s-s-C} as
\begin{align}
\label{eq: OmegaT1}
\text{tr}(\mathbf e_i \mathbf e_i^T [\boldsymbol \Lambda_h s_{\infty} + \boldsymbol\Sigma_{\epsilon}]) \text{tr}(\tilde{\mathbf W}\mathbf e_i \mathbf e_i^T\tilde{\mathbf W}^T) &= \tilde{\mathbf w}^T\boldsymbol{\Omega}_{i}^{\mathrm{T}(0)}\tilde{ \mathbf w}, \\
&= \mathbf w^T\boldsymbol{\Omega}_{i}^{\mathrm{T}(1)}\mathbf w,
\end{align}
where the $(\tilde{k}, \tilde{l})$th entry of $\boldsymbol{\Omega}_{i}^{\mathrm{T}(0)}$ is also obtained according to \eqref{eq: D_w}, namely, $[\boldsymbol{\Omega}_{i}^{\mathrm{T}(0)}]_{\tilde{k} \tilde{l}} = \text{tr}(\mathbf e_i \mathbf e_i^T [\boldsymbol \Lambda_h s_{\infty} + \boldsymbol\Sigma_{\epsilon}])[\mathbf e_i \mathbf e_i^T]_{n_{\tilde{k}} n_{\tilde{l}}}$, here $\tilde{k}, \tilde{l} \in [L-M]$ and $\boldsymbol{\Omega}_{i}^{\mathrm{T}(1)} = \mathbf{F}^T\boldsymbol{\Omega}_{i}^{\mathrm{T}(0)}\mathbf{F}$.

Upon defining $\boldsymbol \Omega_{\mathrm C, i} = \boldsymbol{\Omega}_{i}^{\mathrm{T}(1)}$ as the characterization of expected transmission cost consumed for inter-sensor communication, and $\boldsymbol \Omega_{\mathrm T, i} = \boldsymbol{\Omega}_{i}^{\mathrm{T}(1)} + \boldsymbol{\Omega}_{i}^{\mathrm{T}(2)}$ as the characterization of both the expected transmission cost consumed for inter-sensor communication and sensor-to-FC, we eventually obtain the problem \eqref{eq: prob1_off_equiv1}. In what follows, we summarize the coefficients precisely correspond to the problem \eqref{eq: prob1_off_equiv1}
\begin{align*}
&\boldsymbol{\Omega}_{\mathrm{N}} =  \mathbf G\mathbf h \mathbf h^T \mathbf G^T, \\
&\boldsymbol{\Omega}_{\mathrm{D}}, \\
&\boldsymbol \Omega_{\mathrm C, i} = \boldsymbol{\Omega}_{i}^{\mathrm{T}(1)}, \\
&\boldsymbol \Omega_{\mathrm T, i} = \boldsymbol{\Omega}_{i}^{\mathrm{T}(1)} + \boldsymbol{\Omega}_{i}^{\mathrm{T}(2)}, 
\end{align*}
where $[\boldsymbol{\Omega}_{\mathrm{D}}]_{kl} =  [\boldsymbol \Lambda_g]_{m_k m_l} [\boldsymbol \Lambda_h s_{\infty} + \boldsymbol \Sigma_\epsilon]_{n_k n_l}$, $\boldsymbol{\Omega}_{i}^{\mathrm{T}(1)} = \mathbf{F}^T\boldsymbol{\Omega}_{i}^{\mathrm{T}(0)}\mathbf{F}$, $[\boldsymbol{\Omega}_{i}^{\mathrm{T}(0)}]_{\tilde{k} \tilde{l}} = \text{tr}(\mathbf e_i \mathbf e_i^T [\boldsymbol \Lambda_h s_{\infty} + \boldsymbol\Sigma_{\epsilon}])[\mathbf e_i \mathbf e_i^T]_{n_{\tilde{k}} n_{\tilde{l}}}$ and $[\boldsymbol{\Omega}_{i}^{\mathrm{T}(2)}]_{k l} = [\mathbf e_i \mathbf e_i^T]_{m_k m_l}[\boldsymbol \Lambda_h s_{\infty} + \boldsymbol \Sigma_\epsilon]_{n_k n_l}$, for $k, l \in [L], \tilde{k}, \tilde{l} \in [L-M]$. We note that the matrix $\boldsymbol{\Omega}_{\mathrm{N}}$ is positive semidefinite and of rank one. Since the denominator of $f(\mathbf W)$ in \eqref{eq: prob_off_1} must be positive when the channel noise and collaboration noise are ignored (namely, $\sigma_\varsigma^2 = 0$ and $\sigma_\kappa^2 = 0$), the matrix  $\boldsymbol{\Omega}_{\mathrm{D}}$ is  positive definite. The matrices $\boldsymbol \Omega_{\mathrm C, i}$ and $\boldsymbol \Omega_{\mathrm T, i}$ are also positive definite according  to the definition of transmission cost. \hfill $\blacksquare$

\section{Proof of Proposition\ \ref{prop: SDP_off}}
\label{appendix: Prop3}
By introducing a new variable $\bar{\mathbf W} \in \mathbb R^{(L+1) \times (L+1)}$ together with the constraint $\bar{\mathbf W} = \bar{\mathbf{w}} \bar{\mathbf{w}}^T$, problem \eqref{eq: prob1_off_equiv3} can be reformulated as
\begin{align}
\begin{array}{ll}
\label{eq: prob1_off_equiv4}
\underset{\bar{\mathbf W}}{\text{maximize}} \quad & \text{tr}({\mathbf{Q}_{0}}\bar{\mathbf W}), \\
\vspace{1mm}
\st \quad & \text{tr}({\mathbf{Q}}_{i}^1\bar{\mathbf W}) \leq 0, ~ i \leq M,\\
\vspace{1mm}
& \text{tr}({\mathbf{Q}}_{i}^2\bar{\mathbf W}) \leq 0, ~ M < i \leq N, \\
\vspace{1mm}
& \text{tr}({\mathbf{Q}_{N+1}}\bar{\mathbf W}) \leq 1,~ \text{rank}(\bar{\mathbf W}) = 1,~ \bar{\mathbf W}  \succeq 0,
 \end{array}
\end{align}
where we have used the fact that the constraint  $\text{rank}(\bar{\mathbf W}) = 1$ together with $\bar{\mathbf W} \succeq 0$ is equivalent to the constraint $\bar{\mathbf W} = \bar{\mathbf{w}} \bar{\mathbf{w}}^T$, and $\bar{\mathbf W} \succeq 0$ indicates that $\bar{\mathbf W}$ is positive semidefinite. After dropping the (nonconvex) rank-one constraint, problem \eqref{eq: prob1_off_equiv4} is relaxed to the semidefinite program (SDP) \eqref{eq: prob1_off_equiv5}.

It is clear from \eqref{eq: prob1_off_equiv3}-\eqref{eq: prob1_off_equiv5} that the solution of \eqref{eq: prob1_off_equiv1} is achievable based on the solution of problem \eqref{eq: prob1_off_equiv5} if the latter renders a rank-one solution. Similar to \cite[Theorem\,1]{jiang2014optimal} and \cite[Theorem\,1.4]{huamaizha10},  the rank-one property of $\bar{\mathbf W}^*$ can be explored by studying the KKT conditions of problem \eqref{eq: prob1_off_equiv5}.  Details of the proof are omitted here for the sake of brevity. Since $\bar{\mathbf W}^*$ is of rank one, we can immediately construct \eqref{eq: sol_wstar} as the solution of problem \eqref{eq: prob1_off_equiv1}. \hfill $\blacksquare$


\section{Proof of Lemma\ \ref{lemma: 2}}
\label{appendix: lemma2}
\begin{enumerate}
\item Since $\left | \alpha \right | < 1$, $\theta_k \overset{d}{\rightarrow} \theta_{\infty}$ where $\theta_{\infty}$ is an a.s. finite random variable \cite{grimmett2001probability}. Also, then $\theta_{\infty} \overset{d}{=} \alpha \theta_{\infty} + \tau$, where $\theta_{\infty}$ and $\tau$ are independent, and $\tau$ has the distribution of $\tau_1$ and $\overset{d}{=}$ denotes equality in distribution. Therefore, $\mathbb E[\theta_{\infty}] = \alpha \mathbb E[\theta_{\infty}] + \mathbb E[\tau]$ and hence $\mathbb E[\theta_{\infty}] = \mathbb E[\tau]/(1-\alpha)$. Also, $\mathbb E[\theta_{\infty}^2] = \alpha^2 \mathbb E[\theta_{\infty}^2] + \mathbb E[\tau^2]$, because $\mathbb E[\tau] = 0$. Therefore, $\mathbb E[\theta_{\infty}^2] = \mathbb E[\tau^2]/(1-\alpha^2)$.

From the dynamical equation \eqref{eq: theta_k},
\begin{equation*}
\theta_k = \alpha^k \theta_0 + \tau_k + \alpha \tau_{k-1} + \alpha^2 \tau_{k-2} + \ldots + \alpha^{k-1}\tau_1.
\end{equation*}
Therefore, 
\begin{align*}
\mathbb E[\theta_k] &= \alpha^k \mathbb E[\theta_0] + (1 + \alpha + \ldots + \alpha^{k-1}) \mathbb E[\tau_1] \\ \nonumber
& \to \mathbb E[\tau_1] / (1 - \alpha), \; \text{as} \; k \to \infty \; \text{and} \\
\mathbb E[\theta_k^2] &= \mathbb E[\tau_1^2] (1 + \alpha + \ldots + \alpha^{k-1}) \\
& \to \mathbb E[\tau_1^2] / (1 - \alpha^2).
\end{align*}

\item From \eqref{eq: P_W12} and \eqref{eq: P_W2E}, we have 
\begin{align*}
\mathbb E[T_{k,i}^{(1)}(\mathbf W^*)] =& \left ( \mathbf e_i^T( \mathbb E[\theta_k^2](\mathbf{h}\mathbf{h}^T + \mathbf{\Sigma}_h) + \mathbf{\Sigma}_\epsilon ) \mathbf e_i  \right ) \\ \nonumber
&\times [ \mathbf e_i^T ( \mathbf W^* \circ \tilde{\mathbf I} )^T  ( \mathbf W^* \circ  \tilde{\mathbf I}  )  \mathbf e_i  ], \nonumber \\
\mathbb E[T_{k,i}^{(2)}(\mathbf W^*)] =& \mathbf e_i^T  \mathbf W^* (\mathbb E[\theta_k^2](\mathbf{h}\mathbf{h}^T + \mathbf{\Sigma}_h) + \mathbf{\Sigma}_\epsilon) \\ \nonumber
&\times \mathbf {W^*}^T \mathbf e_i + \sigma_{\kappa}^2. \nonumber
\end{align*}

From the total transmission cost in \eqref{eq: P_W}, we obtain $\mathbb E[T_{k,i}(\mathbf W^*)] \to \mathbb E[T_{i}(\mathbf W^*)]$ since from 1) $\mathbb E[\theta_k^2] \to \mathbb E[\theta_{\infty}^2]$. \hfill $\blacksquare$
\end{enumerate}

\section{Proof of Theorem\ \ref{prop: onlineoffline}}
\label{appendix: Prop4}
In the following, we consider a space $H$ of all real valued random variables on the common underlying probability space, with $\mathbb E[\left | x \right |^2] < \infty$. This is a Hilbert space with inner product $\mathbb E[xy]$ and $\left \| x \right \|^2 = \mathbb E [\left | x \right |^2]$.
Let $\{ y_k \}$ and $\{ y_k^{\prime} \}$ denote the observations received at the FC by using the online sensor collaboration scheme \eqref{eq: policy_given_off} and the offline policy, respectively. The estimates of $\theta_k$ based on $\{ y_k \}$ and $\{ y_k^{\prime} \}$ are given by the projection operator $P_{\mathrm o}$ in the Hilbert space $H$:
\begin{align}
&\hat {\theta}_{k|k} = P_{\mathrm o} (\theta_k | y_1, \ldots, y_k, \hat{\theta}_0), \\
&\hat {\theta}^{\prime}_{k|k} = P_{\mathrm o} (\theta_k | y_1^{\prime}, \ldots, y_k^{\prime}, \hat{\theta}_0). 
\end{align}

Consider the equation
$S_{k+1,i}^{\prime} = (S_{k,i}^{\prime} - T_{k,i}(\mathbf W^*))^{+} + H_{k,i}$ for $i \in [N]$, 
with $S_{0,i}^{\prime}  = S_{0,i}$. 
We have  $\{ H_{k,i}, ~ k \geq 0\}$ and $\{T_{k,i}(\mathbf W^*), k \geq 0\}$, independent, each satisfying SLLN. Also, from Lemma\ \ref{lemma: 2}, $\mathbb E[T_{k,i}(\mathbf W^*)] \to \mathbb E[T_{i}(\mathbf W^*)]$. From offline problem \eqref{eq: prob_off_1}, we have $\mathbb E [T_i(\mathbf W^*)] \leq \mu_i - \eta$. Therefore, since $S_{k+1, i}^{\prime} \geq \sum_{l = 1}^{k}(H_{l,i} - T_{l,i}(\mathbf W^*))$, $S_{k+1,i}^{\prime} \to \infty$ almost surely (a.s.) for all  $i$ as $k \to \infty$. Since the actual energy consumption for sensor $i$ at time $k$ is $T_{k,i}(\mathbf W_k) \leq T_{k,i}(\mathbf W^*)$, we also have $S_{k+1,i} \to \infty$ a.s. for all  $i$.
 Therefore, from \eqref{eq: policy_given_off}, we obtain $T_{k,i}(\mathbf W_k) \to T_i(\mathbf W^*)$ a.s. for all $i$. Accordingly, we obtain that $\mathbf W_{k} \to \mathbf W^*$ a.s. as $k \to \infty$. 
 
 We will use $\| \mathbf x \|$ to denote $\sqrt{\mathbb E [\mathbf x^T \mathbf x]}$ for a random vector $\mathbf x$. 
 
Since $\mathbf W_k \leq \mathbf W^*$, we also obtain $\| \mathbf W_k - \mathbf W^* \| \to 0$ as $k \to \infty$. From \eqref{eq: state_space0}, we have
\begin{align}
 &\| y_k - y_k^{\prime} \|  \leq \\ \nonumber
&\|  \mathbf g_k^T \mathbf W_k \mathbf h_k \theta_k + \mathbf g_k^T\mathbf W_k\boldsymbol \epsilon_k  -  \mathbf g_k^T \mathbf W^* \mathbf h_k  \theta_k - \mathbf g_k^T\mathbf W^*\boldsymbol \epsilon_k \| \\ \nonumber
 &\leq \| \mathbf g_k \| \|  \mathbf W_k -  \mathbf W^* \| \| \boldsymbol \epsilon_k \| + \| \mathbf g_k \| \|  \mathbf W_k -  \mathbf W^* \| \| \mathbf h_k \| \|  \theta_k \|. \nonumber
\end{align}
Since $\| \mathbf g_k \|$, $\| \boldsymbol \epsilon_k \|$, $\| \mathbf h_k\|$ and $\mathbb E [ \theta_\infty^2  ] $ (namely, $s_{\infty}$) are bounded and $\| \mathbf W_k - \mathbf W^* \| \to 0$ and $\left \| \theta_k - \theta_{\infty} \right \| \to 0$, as $k \to \infty$, we obtain
\begin{align}
\| y_k - y_k^{\prime} \|  \to 0, ~  \text{for} ~    k \to \infty. \label{eq: dif_yk}
\end{align}

In what follows, we will show that 
\begin{align}
\|  P_{\mathrm o} (\theta_k | y_1, \ldots, y_k, \hat{\theta}_0) -  P_{\mathrm o} (\theta_k |y_1^{\prime}, \ldots, y_k^{\prime}, \hat{\theta}_0)  \| \to 0,  \label{eq: goal}
\end{align}
for $ k \to \infty$.

From \eqref{eq: theta_k}, we have $\theta_k = \alpha^N \theta_{k-N} + \tau_k + \alpha \tau_{k-1} + \ldots +  \alpha^{N-1} \tau_{k-N+1}$. Therefore, we get
\begin{align}
\label{eq: proj1}
P_{\mathrm o} (\theta_k | y_1, \ldots, y_k, \hat{\theta}_0) &= P_{\mathrm o} ( \alpha^N \theta_{k-N} + \tau_k + \alpha \tau_{k-1} \\ \nonumber
&+ \ldots +  \alpha^{N-1} \tau_{k-N+1} | y_1, \ldots, y_k, \hat{\theta}_0). \nonumber
\end{align}
Since $\{ \tau_k, \ldots, \tau_{k-N+1} \}$ are zero mean and independent of $\{ y_{k-N}, \ldots, y_1, \theta_0 \}$, from \eqref{eq: proj1}, we obtain that
\begin{align}
\label{eq: proj2}
&P_{\mathrm o} (\theta_k | y_1, \ldots, y_k, \hat{\theta}_0) = \alpha^N P_{\mathrm o} (  \theta_{k-N}  | y_1, \ldots, y_k, \hat{\theta}_0) + \\ \nonumber
&P_{\mathrm o} (   \tau_k + \alpha \tau_{k-1} + \ldots +  \alpha^{N-1} \tau_{k-N+1}  | y_k, \ldots, y_{k-N+1} ). \nonumber
\end{align}

For the first term in the RHS of \eqref{eq: proj2}, we note that 
\begin{align}
\alpha^N P_{\mathrm o} (  \theta_{k-N}  | y_1, \ldots, y_k, \hat{\theta}_0) \leq \alpha^N \mathbb E [ \theta_{k-N}^2 ],
\label{eq: proj3}
\end{align}
where we have used the fact that $\|  P_{\mathrm o} (\mathbf x | \mathbf z) \| \leq\| \mathbf x \|$. Moreover, from Lemma\ \ref{lemma: 2}, we obtain that $\mathrm{sup}_{k}E [ \theta_{k-N}^2 ] < \infty$. 
Therefore,  RHS of \eqref{eq: proj3} can be taken arbitrarily small by taking $N$ large enough. In what follows, we keep $N$ fixed.

Now, let us focus on the second term in \eqref{eq: proj2}. For ease of notation, we define
\begin{align}
\bar \tau_k & \Def \tau_k + \alpha \tau_{k-1} + \ldots +  \alpha^{N-1} \tau_{k-N+1}, \\ \nonumber
\bar{\mathbf y}_k & \Def [ y_k, \ldots, y_{k-N+1}]^T,\\ \nonumber
\bar{\mathbf y}_k^{\prime} & \Def [ y_k^{\prime}, \ldots, y_{k-N+1}^{\prime}]^T. \nonumber
\end{align}

From \eqref{eq: dif_yk}, we obtain that
\begin{align}
\| \bar{\mathbf y}_k - \bar{\mathbf y}_k^{\prime}  \| 
\to 0,~ \text{for} ~ k \to \infty.
\end{align}

Then, we have
\begin{align}
\label{eq: Po_key}
&\|  P_{\mathrm o} (\bar \tau_k |\bar{\mathbf y}_k)  - P_{\mathrm o} (\bar \tau_k |\bar{\mathbf y}_k^{\prime}) \| = \left \| \frac{\mathbb E [\bar \tau_k^T  \bar{\mathbf y}_k] \bar{\mathbf y}_k }{ \| \bar{\mathbf y}_k \|^2} - \frac{\mathbb E [\bar \tau_k^T  \bar{\mathbf y}_k^{\prime}] \bar{\mathbf y}_k^{\prime} }{ \| \bar{\mathbf y}_k^{\prime} \|^2} \right \| \nonumber \\
&= \frac{1}{\| \bar{\mathbf y}_k\|^2} \| \mathbb E [\bar \tau_k^T  \bar{\mathbf y}_k] \bar{\mathbf y}_k  - \mathbb E [\bar \tau_k^T  \bar{\mathbf y}_k^{\prime}] \bar{\mathbf y}_k^{\prime} (\| \bar{\mathbf y}_k \|^2/ \| \bar{\mathbf y}_k^{\prime}\|^2)  \|.
\end{align}

In \eqref{eq: Po_key}, we note that $\| \bar{\mathbf y}_k \| \geq \cov(v_k) > 0$ due to \eqref{eq: state_space0}.  
This also holds for $\| \bar{\mathbf y}_k^{\prime} \|$. Moreover, we have $\mathrm{sup}_{k}\| \bar{\mathbf y}_k \| < \infty  $ and $\mathrm{sup}_{k} \| \bar{\mathbf y}_k^{\prime} \| < \infty  $ as $k \to \infty$ due to $\mathrm{sup}_{k} \| \theta_{k} \| < \infty$. Also, we have $(\| \bar{\mathbf y}_k \|^2/ \| \bar{\mathbf y}_k^{\prime}\|^2) \to 1$ as $k \to \infty$ since
$ | \| \bar{\mathbf y}_k \| - \| \bar{\mathbf y}_k^{\prime} \| | \leq \| \bar{\mathbf y}_k - \bar{\mathbf y}_k^{\prime}  \| \to 0 $ as  $k \to \infty$.
Therefore, we obtain
\begin{align}
&\| \mathbb E [\bar \tau_k^T  \bar{\mathbf y}_k] \bar{\mathbf y}_k  - \mathbb E [\bar \tau_k^T  \bar{\mathbf y}_k^{\prime}] \bar{\mathbf y}_k^{\prime} (\| \bar{\mathbf y}_k \|^2/ \| \bar{\mathbf y}_k^{\prime}\|^2)  \| \\ \nonumber
& \leq \| \mathbb E [\bar \tau_k^T  \bar{\mathbf y}_k] \bar{\mathbf y}_k  - \mathbb E [\bar \tau_k^T  \bar{\mathbf y}_k] \bar{\mathbf y}_k^{\prime}  \| \nonumber + \\ \nonumber
&\| \mathbb E [\bar \tau_k^T  \bar{\mathbf y}_k] \bar{\mathbf y}_k^{\prime}  - \mathbb E [\bar \tau_k^T  \bar{\mathbf y}_k^{\prime}] \bar{\mathbf y}_k^{\prime} (\| \bar{\mathbf y}_k \|^2/ \| \bar{\mathbf y}_k^{\prime}\|^2)   \| , \nonumber
\end{align}
where
\begin{align*}
&\| \mathbb E [\bar \tau_k^T  \bar{\mathbf y}_k] \bar{\mathbf y}_k  - \mathbb E [\bar \tau_k^T  \bar{\mathbf y}_k] \bar{\mathbf y}_k^{\prime}  \| \leq \| \mathbb E [\bar \tau_k^T  \bar{\mathbf y}_k] \|  \|  \bar{\mathbf y}_k- \bar{\mathbf y}_k^{\prime}\| \\ \nonumber
&\leq \|  \bar \tau_k \| \|  \bar{\mathbf y}_k] \|  \|  \bar{\mathbf y}_k- \bar{\mathbf y}_k^{\prime}\|  \to 0 ,~ \text{as} ~ k \to \infty \nonumber
\end{align*}
and 
\begin{align*}
&\| \mathbb E [\bar \tau_k^T  \bar{\mathbf y}_k] \bar{\mathbf y}_k^{\prime}  - \mathbb E [\bar \tau_k^T  \bar{\mathbf y}_k^{\prime}] \bar{\mathbf y}_k^{\prime} (\| \bar{\mathbf y}_k \|^2/ \| \bar{\mathbf y}_k^{\prime}\|^2)   \| \\ \nonumber
&\leq \| \bar{\mathbf y}_k^{\prime} \| \| \mathbb E [\bar \tau_k^T  \bar{\mathbf y}_k] -  \mathbb E [\bar \tau_k^T  \bar{\mathbf y}_k^{\prime} (\| \bar{\mathbf y}_k \|^2/ \| \bar{\mathbf y}_k^{\prime}\|^2) ] \|\nonumber \\ \nonumber
& \leq  \| \bar{\mathbf y}_k^{\prime} \|  \| \bar \tau_k \|  \|  \bar{\mathbf y}_k -    \bar{\mathbf y}_k^{\prime} (\| \bar{\mathbf y}_k \|^2/ \| \bar{\mathbf y}_k^{\prime}\|^2)\| \to 0,~ \text{as} ~ k \to \infty. \nonumber
\end{align*}

As a result, RHS of \eqref{eq: Po_key} converges to $0$ as $k \to \infty$. Together with \eqref{eq: proj3}, we have completed the proof of  \eqref{eq: goal}. Therefore, we obtain the result in \eqref{eq: PD}. \hfill $\blacksquare$

\section{Proof of Proposition\ \ref{prop:Wtow}}
\label{appendix: Prop2}
Let $\mathbf w \in \mathbb{R}^L$ be the vector of column-wise \textit{nonzero} entries of $\mathbf W \in \mathbb{R}^{M \times N}$. The one-to-one mapping between $\mathbf w$ and $\mathbf W$ ensures that for any $w_l$, we have a certain pair of indices $(m_l, n_l)$ such that $w_l = W_{m_l n_l}$, where $m_l \in [M], n_l \in [N], l \in [L]$. 

Given $\mathbf b \in \mathbb{R}^M$, we obtain
\begin{equation}
\mathbf b^T\mathbf W = [\mathbf b^T \mathbf W_{\cdot1}  \quad \cdots \quad \mathbf b^T \mathbf W_{\cdot N}],
\end{equation}
where $\mathbf W_{\cdot j}$ is the $j$th column of $\mathbf W$ and $\mathbf b^T\mathbf W_{\cdot j} = \sum_{i=1}^{M}b_{i}W_{ij}$, for $j \in [N]$. Similarly, given $\mathbf B \in \mathbb{R}^{L \times N}$, we have
\begin{equation}
\mathbf w^T\mathbf B = [\mathbf w^T \mathbf B_{\cdot1}  \quad \cdots \quad \mathbf w^T \mathbf B_{\cdot N}],
\end{equation}
where $\mathbf B_{\cdot j}$ is the $j$th column of $\mathbf B$ and $\mathbf w^T\mathbf B_{\cdot j} = \sum_{l=1}^{L}W_{m_l n_l}B_{lj}$, for $j \in [N]$.

Consider the $j$th entry of $\mathbf w^T\mathbf B$, we obtain
\begin{align}
\label{eq: WtowPf}
[\mathbf w^T\mathbf B]_j &= \sum_{l=1}^{L}W_{m_l n_l}B_{lj} = \sum_{l=1, n_l = j}^{L} b_{m_l} W_{m_l j} \\ \nonumber
&= \sum_{m_l=1}^{M}b_{m_l}W_{m_l j} = [\mathbf b^T\mathbf W]_j,
\end{align}
where we have used the facts that $B_{lj} = b_{m_l}$ if $j = n_l$, and $0$ otherwise. Based on \eqref{eq: WtowPf}, we can conclude that $\mathbf w^T \mathbf B = \mathbf b^T\mathbf W$.

We next show identity \eqref{eq: lin12_w2}. Given $\mathbf C \in \mathbb{R}^{M \times M}$ and $\mathbf D \in \mathbb{R}^{N \times N}$, we have 
\begin{equation}
\tr(\mathbf C \mathbf W \mathbf D  \mathbf W^T) = \sum_{i = 1}\mathbf e_i^T \mathbf C \mathbf W \mathbf D  \mathbf W^T \mathbf e_i.
\end{equation}

Let $\mathbf c_i := \mathbf e_i^T \mathbf C$. By applying the identity \eqref{eq: lin12_w1}, we have $\mathbf e_i^T \mathbf C \mathbf W = \mathbf w^T \tilde{\mathbf C}_i$ and $\mathbf W^T\mathbf e_i = \tilde{\mathbf E}_i^T\mathbf w$, where 
\begin{equation*}
\left [ \tilde{\mathbf C}_i \right ]_{ln} = \left \{
\begin{array}{cc}
[\mathbf c_i] _{m_l}, & n = n_l, \\
0, & \text{otherwise}, 
\end{array}
\right. 
\end{equation*}
and 
\begin{equation*}
\left [ \tilde{\mathbf E}_i \right ]_{ln} = \left \{
\begin{array}{cc}
[\mathbf e_i] _{m_l}, & n = n_l, \\
0, & \text{otherwise}, 
\end{array}
\right.
\end{equation*}
for $l \in [L]$ and $n \in [N]$.

Therefore,
\begin{equation}
\label{eq: E1}
\tr(\mathbf C \mathbf W \mathbf D  \mathbf W^T) = \sum_{i = 1} \mathbf w^T \tilde{\mathbf C}_i \mathbf D \tilde{\mathbf E}_i^T\mathbf w = \mathbf w^T (\sum_{i = 1} \tilde{\mathbf C}_i \mathbf D \tilde{\mathbf E}_i^T) \mathbf w.
\end{equation}

Upon defining $\mathbf d_n := \mathbf e_n^T\mathbf D$ and
\begin{equation*}
\left [ \tilde{\mathbf D}_i \right ]_{nk}  = \left \{
\begin{array}{cc}
[\mathbf d_n] _{ n_k }, & m_k = i,   \\
0, & \text{otherwise}, 
\end{array}
\right. 
\end{equation*}
for $n \in [N]$, $l \in [L]$, we obtain
\begin{align}
\label{eq: E2}
\sum_{i = 1} \tilde{\mathbf C}_i \mathbf D \tilde{\mathbf E}_i^T = \sum_{i = 1} \tilde{\mathbf C}_i \tilde{\mathbf D}_i  = \mathbf E,
\end{align} 
where we have used the fact that $E_{l k} = [\mathbf C]_{m_l m_k} [\mathbf D]_{n_l n_k}$
for $n \in [N]$, $k \in [L]$ and $l \in [L]$. Based on \eqref{eq: E2} and \eqref{eq: E1}, we can conclude that $\tr(\mathbf C \mathbf W \mathbf D  \mathbf W^T) = \mathbf w^T \mathbf E \mathbf w$. \hfill $\blacksquare$

\section{Proof of Proposition\ \ref{prop: robust}}
\label{appendix: Proprobust}
From  Proposition\ \ref{prop: SDP_off}, we have the optimal solution $\bar{\mathbf W}^*$ that is of rank one. According to \cite[Theorem\,2.4]{lemon2016low}, we can obtain that $\bar{\mathbf W}^*$ is unique. Moreover, in Proposition\ \ref{prop: SDP_off}, we have shown that the optimal solution of problem (31) is obtained by decomposing the submatrix of $\bar{\mathbf W}^*$ formed by deleting its $(L+1)$st row and column, namely $[\bar{\mathbf W}^*]_L = \mathbf s^*(\mathbf s^*)^T$. Then, if we choose $\left | s^* \right |$, we obtain a unique globally optimal solution $\mathbf W^*$ for the offline problem.

Since the optimizing function of problem (34) is continuous and the feasibility set is compact for a given parameter set, by \cite[Theorem\,9.14]{sundaram1996first}, the unique optimal solution $\bar{\mathbf W}^*$ of problem (34) is a continuous function of the parameters. Hence, the optimal solution $\mathbf W^*$ of problem (31) is also a continuous function.

Next consider the asymptotic mean squared error (MSE) $P_{\infty}$ in (23) of the offline sensor collaboration scheme (and hence also of the online sensor collaboration scheme). Equation (23) is quadratic in $P_{\infty}$ and hence $P_{\infty}$ is a continuous function of its coefficients which are continuous functions of $\mathbf W^*$ and the other parameters. Therefore, $P_{\infty}$ is a continuous function of the system parameters. \hfill $\blacksquare$

\section{Proof of Lemma\ \ref{lemma: 3}}
\label{appendix: lemma3}
In the following, we show the convergence rate of $\left \| \hat{\theta}_{k|k} - \theta_k \right \| \to P_{\infty}$, where $\hat {\theta}_{k|k}$ denotes the estimate using the proposed optimal online sensor collaboration scheme $\mathbf W_k$ and $\theta_k$ is the true value.

According to triangle inequality, we have
\begin{equation}
\left \| \hat{\theta}_{k|k} - \theta_k \right \| \leq \left \| \hat{\theta}_{k|k} - \hat{\theta}_{k|k}^{\prime} \right \| + \left \| \hat{\theta}_{k|k}^{\prime} - \theta_k \right \|,
\end{equation}
where $\hat {\theta}^{\prime}_{k|k}$ denote the estimate when using the offline policy $\mathbf W^*$.

From \cite[Theorem\,2]{Tug81}, $\left \| \hat{\theta}_{k|k}^{\prime} - \theta_k \right \|$ converges to $P_{\infty}$ exponentially. Now, we consider $\left \| \hat{\theta}_{k|k} - \hat{\theta}_{k|k}^{\prime} \right \|$.

From (65) in proof of Theorem 1, for any $N > 0$, we have
\begin{equation*}
\left \| \hat{\theta}_{k|k} - \hat{\theta}_{k|k}^{\prime} \right \| \leq \alpha^N \mathbb E[\theta_{k - N}^2] + \left \| P_o(\bar{\tau}_k | \bar{y}_k) - P_o(\bar{\tau}_k | \bar{y}_k^{\prime}) \right \|.
\end{equation*}
where $\bar \tau_k \Def \tau_k + \alpha \tau_{k-1} + \ldots +  \alpha^{N-1} \tau_{k-N+1}$, 
$\bar{\mathbf y}_k \Def [ y_k, \ldots, y_{k-N+1}]^T$, $\bar{\mathbf y}_k^{\prime} \Def [ y_k^{\prime}, \ldots, y_{k-N+1}^{\prime}]^T$ and $P_o$ is the projection operator which determines the estimate of $\theta_k$.

By taking $N$ large enough, the first term on RHS can be taken arbitrarily small. This also happens at an exponential rate.

From (69) and the inequality below it, we have 
\begin{equation*}
\left \| P_o(\bar{\tau}_k / \bar{y}_k) - P_o(\bar{\tau}_k / \bar{y}_k^{\prime}) \right \| \leq c \left \| \bar{y}_k - \bar{y}_k^{\prime} \right \|
\end{equation*}
for $c$ is an appropriately large constant. Using triangle inequality and Cauchy-Schwarz inequality, we obtain
\begin{equation}
\label{eq:yk1}
\left \| \bar{y}_k - \bar{y}_k^{\prime} \right \| \leq \sum_{j=k-N+1}^{k} \left \| y_j - y_j^{\prime} \right \|, \end{equation}
and 
\begin{align}
\left \| y_j - y_j^{\prime} \right \|^2 \!&= \mathbb E [((\mathbf g_j^T\mathbf W^*\mathbf h_j\theta_j + v_j)\! - \!(\mathbf g_j^T\beta_{j}  \circ \mathbf W^*\mathbf h_j\theta_j + v_j))^2] \\ \nonumber
&\leq \left \| \mathbf g_j \right \|^2 \left \| \mathbf h_j \right \|^2 \left \| \theta_j \right \|^2 \left \| \mathbf W^* - \beta_{j}  \circ \mathbf W^* \right \|^2. \nonumber
\end{align}

Also, $\left \| \mathbf W^* - \beta_{k}  \circ \mathbf W^* \right \| \leq \mathbb E[(1 - \beta_k)^2] \left \| \mathbf W^* \right \|$.
Therefore, from \eqref{eq:yk1}, we have
\begin{equation*}
\left \| \bar{y}_k - \bar{y}_k^{\prime} \right \| \leq \left \| \mathbf g \right \| \left \| \mathbf h \right \| \left \| \mathbf W^* \right \| \sum_{j=k-N+1}^{k} \left \| \theta_j \right \| \mathbb E[(1 - \beta_j)^2].
\end{equation*}

Since $\underset{j}{\text{sup}}\left \| \theta_j \right \| < \infty$ and $N$ is taken a finite fixed number, we then need to show rate of convergence of $\mathbb E[(1 - \beta_k)^2]$. For $\delta_k > 0$, 
\begin{align}
\mathbb E[(1 - \beta_k)^2] &= \mathbb E[(1 - \beta_k)^2 | \beta_k \leq \delta_k] P(\beta_k \leq \delta_k) \\ \nonumber
&+ \mathbb E[(1 - \beta_k)^2 | \beta_k > \delta_k] P(\beta_k > \delta_k) \\ \nonumber
&\leq P(\beta_k \leq \delta_k) + (1 - \delta_k)^2. \nonumber
\end{align}

Take $\delta_k = 1 - \frac{1}{\delta^k}$ for a small $\delta > 0$. Then second term $(1 - \delta_k)^2 = \frac{1}{\delta^{2k}}$ and decays exponentially. Now we consider $P(\beta_k \leq \delta_k)$.
\begin{align*}
&P(\beta_k \leq \delta_k) = P(\beta_k \leq 1 - \frac{1}{\delta^k}) \\
&= P(S_{k,i} \leq (1 - \frac{1}{\delta^k})^2T_{k,i}(\mathbf W^*), i = 1, \ldots, N) \\
&\leq \sum_{i=1}^{N}P\left [S_{k,i} \leq (1 - \frac{1}{\delta^k})^2T_{k,i}(\mathbf W^*)\right ]  \\
&\leq \sum_{i=1}^{N}P\left [\sum_{j=1}^{k}H_{j,i} - \sum_{j=1}^{k}T_{j,i}(\mathbf W^*) \leq (1 - \frac{1}{\delta^k})^2 T_{k,i}(\mathbf W^*)\right ]  \\
&\leq \sum_{i=1}^{N}P\left [\sum_{j=1}^{k}(H_{j,i} - T_{j,i}(\mathbf W^*)) \leq T_{k,i}(\mathbf W^*)\right ] \\
&= \sum_{i=1}^{N}\!\!P\!\!\left [\frac{1}{k}\!\sum_{j=1}^{k} (H_{j,i} - T_{j,i}(\mathbf W^*))\!\! \leq \!\!\frac{T_{k,i}(\mathbf W^*)}{k}; \!\! \frac{T_{k,i}(\mathbf W^*)}{k} \!\! \leq\! \varpi \!\right ] + \\
& \sum_{i=1}^{N}\!P\!\!\left [\frac{1}{k}\sum_{j=1}^{k}\!(H_{j,i} - T_{j,i}(\mathbf W^*)) \!\! \leq \!\!\frac{T_{k,i}(\mathbf W^*)}{k}; \! \frac{T_{k,i}(\mathbf W^*)}{k} \! > \! \varpi \right ] 
\end{align*}
\begin{align*}
&\leq \sum_{i=1}^{N}\!\!P\!\!\left [\frac{1}{k}\sum_{j=1}^{k} (H_{j,i} - T_{j,i}(\mathbf W^*)) \leq \varpi) + P(\frac{T_{k,i}(\mathbf W^*)}{k}\! > \!\varpi) \right ],
\end{align*}
where we choose $\varpi < \eta$. Since $\mathbb E[T_i(\mathbf W^*)] \leq \mathbb E[H_i] - \eta$, then $\mathbb E[H_i] - \mathbb E[T_i(\mathbf W^*)] \geq \eta$. \hfill $\blacksquare$

\bibliographystyle{IEEEbib}
\bibliography{journal_shan,reff_shan}

\begin{thebibliography}{10}

\bibitem{yick2008wireless}
Jennifer Yick, Biswanath Mukherjee, and Dipak Ghosal,
\newblock ``Wireless sensor network survey,''
\newblock {\em Computer networks}, vol. 52, no. 12, pp. 2292--2330, 2008.

\bibitem{cheng2012survey}
Long Cheng, Chengdong Wu, Yunzhou Zhang, Hao Wu, Mengxin Li, and Carsten Maple,
\newblock ``A survey of localization in wireless sensor network,''
\newblock {\em International Journal of Distributed Sensor Networks}, 2012.

\bibitem{he2006achieving}
Tian He, Pascal Vicaire, Ting Yan, Liqian Luo, Lin Gu, Gang Zhou, Radu Stoleru,
  Qing Cao, John~A Stankovic, and Tarek Abdelzaher,
\newblock ``Achieving real-time target tracking using wireless sensor
  networks,''
\newblock in {\em Real-Time and Embedded Technology and Applications Symposium,
  2006. Proceedings of the 12th IEEE}. IEEE, 2006, pp. 37--48.

\bibitem{olfati2007consensus}
Reza Olfati-Saber, J~Alex Fax, and Richard~M Murray,
\newblock ``Consensus and cooperation in networked multi-agent systems,''
\newblock {\em Proceedings of the IEEE}, vol. 95, no. 1, pp. 215--233, 2007.

\bibitem{kar2013linear}
Swarnendu Kar and Pramod~K Varshney,
\newblock ``Linear coherent estimation with spatial collaboration,''
\newblock {\em IEEE Transactions on Information Theory}, vol. 59, no. 6, pp.
  3532--3553, 2013.

\bibitem{cuixiagolluopoo07}
S.~Cui, J.-J. Xiao, A.~J. Goldsmith, Z.-Q. Luo, and H.~V. Poor,
\newblock ``Estimation diversity and energy efficiency in distributed
  sensing,''
\newblock {\em IEEE Transactions on Signal Processing}, vol. 55, no. 9, pp.
  4683--4695, 2007.

\bibitem{xiao2008linear}
Jin-Jun Xiao, Shuguang Cui, Zhi-Quan Luo, and Andrea~J Goldsmith,
\newblock ``Linear coherent decentralized estimation,''
\newblock {\em IEEE Transactions on Signal Processing}, vol. 56, no. 2, pp.
  757--770, 2008.

\bibitem{knorn2015distortion}
Steffi Knorn, Subhrakanti Dey, Anders Ahl{\'e}n, and Daniel~E Quevedo,
\newblock ``Distortion minimization in multi-sensor estimation using energy
  harvesting and energy sharing.,''
\newblock {\em IEEE Trans. Signal Processing}, vol. 63, no. 11, pp. 2848--2863,
  2015.

\bibitem{leong2011asymptotics}
Alex~S Leong, Subhrakanti Dey, and Jamie~S Evans,
\newblock ``Asymptotics and power allocation for state estimation over fading
  channels,''
\newblock {\em IEEE Transactions on Aerospace and Electronic Systems}, vol. 47,
  no. 1, pp. 611--633, 2011.

\bibitem{jiang2014optimal}
Feng Jiang, Jie Chen, and A~Lee Swindlehurst,
\newblock ``Optimal power allocation for parameter tracking in a distributed
  amplify-and-forward sensor network,''
\newblock {\em IEEE Transactions on Signal Processing}, vol. 62, no. 9, pp.
  2200--2211, 2014.

\bibitem{fang2009power}
Jun Fang and Hongbin Li,
\newblock ``Power constrained distributed estimation with correlated sensor
  data,''
\newblock {\em IEEE Transactions on Signal Processing}, vol. 57, no. 8, pp.
  3292--3297, 2009.

\bibitem{thatte2008sensor}
Gautam Thatte and Urbashi Mitra,
\newblock ``Sensor selection and power allocation for distributed estimation in
  sensor networks: Beyond the star topology,''
\newblock {\em IEEE Transactions on Signal Processing}, vol. 56, no. 7, pp.
  2649--2661, 2008.

\bibitem{liu2014sparsity}
Sijia Liu, Swarnendu Kar, Makan Fardad, and Pramod~K Varshney,
\newblock ``Sparsity-aware sensor collaboration for linear coherent
  estimation,''
\newblock {\em IEEE Transactions on Signal Processing}, vol. 63, no. 10, pp.
  2582--2596, 2014.

\bibitem{liu2015optimal}
Sijia Liu, Swarnendu Kar, Makan Fardad, and Pramod~K Varshney,
\newblock ``On optimal sensor collaboration for distributed estimation with
  individual power constraints,''
\newblock in {\em Signals, Systems and Computers, 2015 49th Asilomar Conference
  on}. IEEE, 2015, pp. 571--575.

\bibitem{liu2016optimized}
Sijia Liu, Swarnendu Kar, Makan Fardad, and Pramod~K Varshney,
\newblock ``Optimized sensor collaboration for estimation of temporally
  correlated parameters,''
\newblock {\em IEEE Transactions on Signal Processing}, vol. 64, no. 24, pp.
  6613--6626, 2016.

\bibitem{priya2009energy}
Shashank Priya and Daniel~J Inman,
\newblock {\em Energy harvesting technologies}, vol.~21,
\newblock Springer, 2009.

\bibitem{zhao2013optimal}
Yu~Zhao, Biao Chen, and Rui Zhang,
\newblock ``Optimal power allocation for an energy harvesting estimation
  system,''
\newblock in {\em Acoustics, Speech and Signal Processing (ICASSP), 2013 IEEE
  International Conference on}. IEEE, 2013, pp. 4549--4553.

\bibitem{huang2013power}
Chuan Huang, Yang Zhou, Tao Jiang, Ping Zhang, and Shuguang Cui,
\newblock ``Power allocation for joint estimation with energy harvesting
  constraints,''
\newblock in {\em Acoustics, Speech and Signal Processing (ICASSP), 2013 IEEE
  International Conference on}. IEEE, 2013, pp. 4804--4808.

\bibitem{nayyar2013optimal}
Ashutosh Nayyar, Tamer Ba{\c{s}}ar, Demosthenis Teneketzis, and Venugopal~V
  Veeravalli,
\newblock ``Optimal strategies for communication and remote estimation with an
  energy harvesting sensor,''
\newblock {\em IEEE Transactions on Automatic Control}, vol. 58, no. 9, pp.
  2246--2260, 2013.

\bibitem{liu2016optimal}
Sijia Liu, Yanzhi Wang, Makan Fardad, and Pramod~K Varshney,
\newblock ``Optimal energy allocation and storage control for distributed
  estimation with sensor collaboration,''
\newblock in {\em Information Science and Systems (CISS), 2016 Annual
  Conference on}. IEEE, 2016, pp. 42--47.

\bibitem{liutowards}
Sijia Liu, Vinod Sharma, and Pramod~K Varshney,
\newblock ``Towards an online energy allocation policy for distributed
  estimation with sensor collaboration using energy harvesting sensors,''
\newblock in {\em Signal and Information Processing (GlobalSIP), 2016 Annual
  Conference on}. IEEE, 2016.

\bibitem{mudumbai2009distributed}
Raghuraman Mudumbai, D~Richard~Brown Iii, Upamanyu Madhow, and H~Vincent Poor,
\newblock ``Distributed transmit beamforming: challenges and recent progress,''
\newblock {\em IEEE Communications Magazine}, vol. 47, no. 2, pp. 102--110,
  2009.

\bibitem{li2007distributed}
Wenjun Li and Huaiyu Dai,
\newblock ``Distributed detection in wireless sensor networks using a multiple
  access channel,''
\newblock {\em IEEE Transactions on Signal Processing}, vol. 55, no. 3, pp.
  822--833, 2007.

\bibitem{bucklew2008convergence}
James~A Bucklew and William~A Sethares,
\newblock ``Convergence of a class of decentralized beamforming algorithms,''
\newblock {\em IEEE Transactions on Signal Processing}, vol. 56, no. 6, pp.
  2280--2288, 2008.

\bibitem{meyn2012markov}
Sean~P Meyn and Richard~L Tweedie,
\newblock {\em Markov chains and stochastic stability},
\newblock Springer Science \& Business Media, 2012.

\bibitem{sharma2010optimal}
Vinod Sharma, Utpal Mukherji, Vinay Joseph, and Shrey Gupta,
\newblock ``Optimal energy management policies for energy harvesting sensor
  nodes,''
\newblock {\em IEEE Transactions on Wireless Communications}, vol. 9, no. 4,
  2010.

\bibitem{rajesh2014capacity}
Ramachandran Rajesh, Vinod Sharma, and Pramod Viswanath,
\newblock ``Capacity of gaussian channels with energy harvesting and processing
  cost,''
\newblock {\em IEEE Transactions on Information Theory}, vol. 60, no. 5, pp.
  2563--2575, 2014.

\bibitem{michelusi2012optimal}
Nicolo Michelusi, Kostas Stamatiou, and Michele Zorzi,
\newblock ``On optimal transmission policies for energy harvesting devices,''
\newblock in {\em Information Theory and Applications Workshop (ITA), 2012}.
  IEEE, 2012, pp. 249--254.

\bibitem{michelusi2013optimal}
Nicolo Michelusi and Michele Zorzi,
\newblock ``Optimal random multiaccess in energy harvesting wireless sensor
  networks,''
\newblock in {\em Communications Workshops (ICC), 2013 IEEE International
  Conference on}. IEEE, 2013, pp. 463--468.

\bibitem{aprem2013transmit}
Anup Aprem, Chandra~R Murthy, and Neelesh~B Mehta,
\newblock ``Transmit power control policies for energy harvesting sensors with
  retransmissions,''
\newblock {\em IEEE Journal of Selected Topics in Signal Processing}, vol. 7,
  no. 5, pp. 895--906, 2013.

\bibitem{kansal2007power}
Aman Kansal, Jason Hsu, Sadaf Zahedi, and Mani~B Srivastava,
\newblock ``Power management in energy harvesting sensor networks,''
\newblock {\em ACM Transactions on Embedded Computing Systems (TECS)}, vol. 6,
  no. 4, pp. 32, 2007.

\bibitem{mao2013energy}
Yuyi Mao, Guanding Yu, and Caijun Zhong,
\newblock ``Energy consumption analysis of energy harvesting systems with power
  grid,''
\newblock {\em IEEE Wireless Communications Letters}, vol. 2, no. 6, pp.
  611--614, 2013.

\bibitem{feinberg2012handbook}
Eugene~A Feinberg and Adam Shwartz,
\newblock {\em Handbook of Markov decision processes: methods and
  applications}, vol.~40,
\newblock Springer Science \& Business Media, 2012.

\bibitem{altman1999constrained}
Eitan Altman,
\newblock {\em Constrained Markov decision processes}, vol.~7,
\newblock CRC Press, 1999.

\bibitem{donoho1982breakdown}
David~L Donoho,
\newblock ``Breakdown properties of multivariate location estimators,''
\newblock Tech. {R}ep., Technical report, Harvard University, Boston. URL
  http://www-stat. stanford. edu/\~{} donoho/Reports/Oldies/BPMLE. pdf, 1982.

\bibitem{rajsatsri71}
P.~K. Rajasekaran, N.~Satyanarayana, and M.~D. Srinath,
\newblock ``Optimum linear estimation of stochastic signals in the presence of
  multiplicative noise,''
\newblock {\em IEEE Transactions on Aerospace and Electronic Systems}, vol. 7,
  no. 3, pp. 462--468, May 1971.

\bibitem{Tug81}
J.~Tugnait,
\newblock ``Stability of optimum linear estimators of stochastic signals in
  white multiplicative noise,''
\newblock {\em IEEE Transactions on Automatic Control}, vol. 26, no. 3, pp.
  757--761, Jun 1981.

\bibitem{huamaizha10}
Y.~Huang, D.~A. Maio, and S.~Zhang,
\newblock ``Semidefinite programming, matrix decomposition, and radar code
  design,''
\newblock in {\em Convex Optimization in Signal Processing and Communications},
  pp. 192--228. Cambridge University Press, New York, 2010.

\bibitem{iutzeler2012analysis}
Franck Iutzeler, Philippe Ciblat, and J{\'e}r{\'e}mie Jakubowicz,
\newblock ``Analysis of max-consensus algorithms in wireless channels,''
\newblock {\em IEEE Transactions on Signal Processing}, vol. 60, no. 11, pp.
  6103--6107, 2012.

\bibitem{adamczak2015exponential}
Rados{\l}aw Adamczak and Witold Bednorz,
\newblock ``Exponential concentration inequalities for additive functionals of
  markov chains,''
\newblock {\em ESAIM: Probability and Statistics}, vol. 19, pp. 440--481, 2015.

\bibitem{freris2010fundamentals}
Nikolaos~M Freris, Hemant Kowshik, and PR~Kumar,
\newblock ``Fundamentals of large sensor networks: Connectivity, capacity,
  clocks, and computation,''
\newblock {\em Proceedings of the IEEE}, vol. 98, no. 11, pp. 1828--1846, 2010.

\bibitem{grimmett2001probability}
Geoffrey Grimmett and David Stirzaker,
\newblock {\em Probability and random processes},
\newblock Oxford university press, 2001.

\bibitem{lemon2016low}
Alex Lemon, Anthony Man-Cho So, Yinyu Ye, et~al.,
\newblock ``Low-rank semidefinite programming: Theory and applications,''
\newblock {\em Foundations and Trends{\textregistered} in Optimization}, vol.
  2, no. 1-2, pp. 1--156, 2016.

\bibitem{sundaram1996first}
Rangarajan~K Sundaram,
\newblock {\em A first course in optimization theory},
\newblock Cambridge university press, 1996.

\end{thebibliography}

\end{document}